\documentclass[useAMS,usenatbib,usegraphicx]{mn2e}

\usepackage{journals}
\usepackage{url}
\usepackage{multirow}
\newcommand{\thickhline}{\noalign{\hrule height 0.8pt}}
\usepackage{threeparttable}

\title[NIR/MIR constraints on 7 XRBs]{Sample of optically unidentified X-ray binaries in the Galactic bulge. Constraints on the physical nature from infrared photometric surveys}
\author[Zolotukhin, and Revnivtsev]{Ivan Yu. Zolotukhin$^{1,2}$\thanks{E-mail:
ivan.zolotukhin@irap.omp.eu (IZ)}, and Mikhail G.
Revnivtsev$^{3}$\\
$^{1}$CNRS, IRAP, 9 avenue du Colonel Roche, BP 44346, F-31028 Toulouse Cedex 4, France\\
$^{2}$Sternberg Astronomical Institute, Moscow State University, Universitetskij pr., 13, 119992, Moscow, Russia\\
$^{3}$Space Research Institute, Russian Academy of Sciences, Profsoyuznaya 84/32, 117997 Moscow, Russia\\
}
\begin{document}

\date{Accepted 2014 October 21. Received 2014 October 20; in original form 2013 August 19}

\pagerange{\pageref{firstpage}--\pageref{lastpage}} \pubyear{2014}

\maketitle

\label{firstpage}

\begin{abstract}
We report on the archival near-infrared and mid-infrared observations of 7 persistent X-ray sources situated in the Galactic bulge using data from the UKIRT Infrared Deep Sky Survey (UKIDSS), {\it Spitzer} Galactic Legacy Infrared Mid-Plane Survey Extraordinaire (GLIMPSE) and the Wide-field Infrared Survey Explorer ({\it WISE}) all-sky survey. We were able to successfully identify, or provide upper flux limits for the systems SAX~J1747.0$-$2853, IGR~J17464$-$2811, AX~J1754.2$-$2754, IGR~J17597$-$2201, IGR~J18134$-$1636, IGR~J18256$-$1035, Ser~X$-$1 and constrain the nature of these systems. In the case of IGR~J17597$-$2201 we present arguments that the source accretes matter from the stellar wind rather than via Roche lobe overflow of the secondary. We suggest that, at its X-ray luminosity of $10^{34-35}$~erg~s$^{-1}$, we are probing the poorly known class of wind-fed low-mass X-ray binaries (LMXBs).
\end{abstract}


\section{Introduction}

X-ray surveys of the sky, performed over the last decade, have provided a lot of information about populations of sources in our Galaxy \citep[see e.g.][]{grimm02,gilfanov04,sazonov06}. The hunt for the complete samples of sources is motivated by the desire of better statistics to test theoretical predictions of binary system evolution, which in turn helps to measure physical effects, unreachable by any other methods.

Since our Galaxy has been well-studied in the X-ray bandpass \cite[see e.g. survey of the sky of {\it Uhuru} satellite by][]{forman78}, a large fraction of the brightest sources was securely identified and a lot of additional properties were catalogued \cite[see e.g][]{liu07}. Studies of these samples of known sources allows one to connect the intrinsic properties of the population with features of their observational appearance. For example, it was shown that the number of low-mass X-ray binaries traces the stellar density \citep{grimm02,gilfanov04}, while number of high-mass X-ray binaries in galaxies traces their total star formation rate \citep{grimm03,ranalli03}; the spatial distribution of high-mass X-ray binaries and their correspondence with the star formation regions makes it possible to constrain their natal kicks \citep{bodaghee12,coleiro13}; the distribution of low-mass X-ray binaries over the luminosity demonstrates features: 1) corresponding to the Eddington luminosity limit of neutron stars \citep{gilfanov04}, 2) corresponding to change of the donor stars in binaries \citep{revnivtsev11}, etc.

More recent X-ray surveys of the Galaxy revealed fainter Galactic X-ray sources which require optical identification. 
Typical sensitivity of the latest surveys is $\simeq 5\times10^{-12}$~erg~s$^{-1}$~cm$^{-2}$ in 17--60~keV energy range \citep[see e.g.][]{krivonos12}, covering a large portion of the Galactic disc, which corresponds to luminosities $L_{\rm X}\simeq 10^{33}$--$10^{35}$~erg~s$^{-1}$ at distances 2-10~kpc.

This luminosity range has peculiarities. 
X-ray binaries with such luminosities, when mass transfer is expected to happen through the Roche lobe overflow of the secondary, can not have stable mass transfer in their accretion discs due to thermal instability \citep[see e.g.][]{meyer81,dubus99,lasota01} and thus should be transient unless they are extremely compact and have orbital periods less than $\simeq$hour. 
In the case of compact systems they have hotter discs with temperatures above the threshold below which a disc becomes thermally unstable \citep{dubus01}.
On the other hand sources might have such low luminosities if they accrete not through the Roche lobe overflow of the secondary, but due to capture of its stellar wind.
Sources of the latter type are historically called symbiotic stars and they are known to be very rare \citep{masetti07}. 
Another peculiarity of this luminosity range is that when material is accreting on to old neutron stars with a surface magnetic field of $10^{8}$~G, at typical mass accretion rates, a so-called `propeller' effect \citep{illarionov75,campana95} is expected, which we suspect reduces the luminosity of the source to below our ability to detect. 
Therefore the systematic search and identification of faint LMXB candidates, to assess their frequency, can give a key insight into the underlying accretion physics in these systems.

In this paper which continue our optical identification efforts started in \citet{zolotukhin10a,zolotukhin10b}. We combine data from the UKIRT Infrared Deep Sky Survey (UKIDSS) \citep{lawrence07}, {\it Spitzer} Galactic Legacy Infrared Mid-Plane Survey Extraordinaire (GLIMPSE) \citep{benjamin03,churchwell09}, and Wide-field Infrared Survey Explorer ({\it WISE}) all-sky survey \citep{wright10} on a set of assumed X-ray binaries, mainly LMXB candidates hosting neutron stars, and present estimates of their NIR brightness, and use these to infer their underlying nature.

\section{NIR Data}
\label{sec_data}

In this study we used several NIR data collections, accessing them through available Virtual Observatory interfaces, such as the TOPCAT tool \citep{taylor05} and the VizieR ConeSearch service for GLIMPSE and {\it WISE} all-sky catalogues, and the WFCAM Science Archive\footnote{\url{http://surveys.roe.ac.uk/wsa/}} for UKIDSS images. Out of several surveys comprising the UKIDSS we used the Galactic Plane Survey and its Data Release 7 (UKIDSS GPS DR7) and in one case of Ser~X$-$1 additionaly Data Release 9. 

We employed our own processing technique for the UKIDSS imaging data instead of using available tabular data in order to detect and measure fainter objects within X-ray positional uncertainties of our objects, which are not present in the public catalogue. 
We considered as potential NIR counterparts only sources situated within $1\sigma$ uncertainty of X-ray coordinates.
We performed PSF photometry measurements of 2~arcmin by 2~arcmin cutouts of UKIDSS science frames using {\sc IRAF daophot} package. 
The calibration of instrumental magnitudes was achieved by comparing our 2~arcsec aperture magnitudes with UKIDSS catalogue magnitudes measured in the same aperture.
We then applied constructed calibration to the PSF magnitudes and hence our photometry results are given in the intrinsic UKIDSS photometric system, i.e. Vega system.
Typically we were able to perform measurements on 1 magnitude fainter stars than those available in the catalogue.
Upper limits of UKIDSS images were estimated using magnitudes of the faintest stars detected at $3\sigma$ level above the background.
Our astrometric calibration is based on the default UKIDSS astrometric solution embedded in UKIDSS image files.
All our astrometric uncertainties include systematic uncertainty of the UKIDSS system with respect to the International Celestial Reference Frame (ICRS).
More details on our photometric and astrometric procedure is given in \citet{zolotukhin10b}.

GLIMPSE data were taken from the official catalogue release, which we accessed through the CDS VizieR service.
When dealing with GLIMPSE data we took into account its astrometric uncertainty which amounts to 0.3~arcsec for each coordinate.
We accessed {\it WISE} all-sky catalogue through the CDS VizieR service as well.
Astrometric uncertainty of {\it WISE} catalogue is known to be 0.4~arcsec for each coordinate.
We list all photometric bands and surveys we used in Table~\ref{tab_bands}.

In each field we computed chance superposition probability which represents the probability of accidental background star coincidence with X-ray $1\sigma$ positional uncertainty. 
It is used to quantify local star density on the $K$ band image in the vicinity of objects and we calculate it as follows: $P = 1 - e^{- S \times \rho}$, where $S$ is an X-ray error circle area and $\rho$ is an average surface density of the sources we detected in 2~arcmin field around given position.

Using our NIR photometry we plotted NIR colour--magnitude diagrams (CMDs) for all stars detected in our fields around X-ray coordinates to test if possible counterpart displays similar colour properties as normal stars. 
For convenience we represent such diagrams as 2D maps of star density in the bins of colour--magnitude space.
The absolute densities in selected bins are not important and this is the reason why we do not give this scale.
The purpose of CMDs is to compare the NIR colour of majority of field stars with the colour of possible counterpart.
As in this sample of X-ray sources we are looking in the direction of the Galactic bulge, a significant fraction of field stars belongs to the bulge population of red giants and subgiants with similar colours residing at the same distance and subject to the same reddening.
If a positional counterpart candidate exhibits same colour as the majority of the field stars, it is unlikely to be associated with the X-ray object.
The nature of the NIR emission in X-ray sources differs from the one in the field stars and hence NIR colours must be different too, unless the interstellar reddening compensates for exact difference which is highly unlikely.
We discuss individual CMDs in each object's section separately.

\begin{table}
\centering
\caption{Bands and surveys used in this study.}
\label{tab_bands}
\begin{tabular}{lccc}
\thickhline
Survey & Band & $\lambda$ & $\Delta \lambda$ \\
\hline
 & & ($\mu$m) & ($\mu$m) \\
\thickhline

UKIDSS & $J$ & 1.25 & 0.079 \\
UKIDSS & $H$ & 1.63 & 0.146 \\
UKIDSS & $K$ & 2.20 & 0.177 \\
{\it Spitzer} & 3.6~$\mu$m & 3.53 & 0.37 \\
{\it Spitzer} & 4.5~$\mu$m & 4.47 & 0.50 \\
{\it Spitzer} & 5.8~$\mu$m & 5.68 & 0.693 \\
{\it Spitzer} & 8.0~$\mu$m & 7.75 & 1.409 \\
{\it WISE} & W1 & 3.38 & 0.34 \\
{\it WISE} & W2 & 4.63 & 0.525 \\
{\it WISE} & W3 & 12.33 & 3.228 \\
{\it WISE} & W4 & 22.25 & 1.973 \\

\thickhline
\end{tabular}
\end{table}

\begin{table*}
\centering
\caption{Observation log for the UKIDSS data used in this study.}
\label{tab_obs}
\begin{tabular}{lccccc}
\thickhline
Field & Obs. start & Filter & Exp. time & Seeing & Mag. limit \\
\hline
 & UTC & & (s) & (arcsec) & Vega mag.\\
\thickhline

SAX J1747.0$-$2853 & 2006-07-18 09:04:42 & $J$ & 10 & 0.8 & 20.2 \\
 & 2006-07-18 09:12:48 & $H$ & 10 & 0.8 & 19.0 \\
 & 2006-07-18 09:20:50 & $K$ & 5 & 1.0 & 17.8 \\
IGR J17464$-$2811  & 2006-07-18 09:06:35 & $J$ & 10 & 0.9 & 20.0 \\
  & 2006-07-18 09:14:40 & $H$ & 10 & 0.9 & 18.7 \\
  & 2006-07-18 09:22:01 & $K$ & 5 & 0.9 & 17.5 \\
AX J1754.2$-$2754  & 2007-05-03 12:51:07 & $J$ & 10 & 0.8 & 19.7 \\
  & 2007-05-03 12:58:53 & $H$ & 10 & 0.9 & 18.7 \\
  & 2007-05-03 13:04:41 & $K$ & 5 & 0.9 & 18.0 \\
IGR J17597$-$2201  & 2006-07-26 08:06:33 & $J$ & 10 & 1.0 & 20.0 \\
  & 2006-07-26 08:14:39 & $H$ & 10 & 1.0 & 19.2 \\
  & 2006-07-26 08:22:08 & $K$ & 5 & 1.1 & 18.1 \\
IGR J18134$-$1636  & 2006-07-23 10:38:22 & $J$ & 10 & 1.0 & 20.0 \\
  & 2006-07-23 10:46:27 & $H$ & 10 & 1.0 & 19.0 \\
  & 2006-07-23 10:54:39 & $K$ & 5 & 0.9 & 18.0 \\
IGR J18256$-$1035  & 2006-06-04 11:54:58 & $J$ & 10 & 0.8 & 20.4 \\
  & 2006-06-04 12:03:27 & $H$ & 10 & 0.8 & 19.5 \\
  & 2006-06-04 12:11:56 & $K$ & 5 & 0.7 & 18.4 \\
Ser X$-$1                   & 2007-05-17 13:54:17 & $K$ & 5 & 0.7 & 18.8 \\
 & 2010-05-12 13:23:23 & $J$ & 10 & 0.8 & 20.4 \\
  & 2010-05-12 13:30:52 & $H$ & 10 & 0.8 & 19.5 \\
  & 2010-05-12 13:36:28 & $K$ & 5 & 0.8 & 18.7 \\

\thickhline
\end{tabular}
\end{table*}

\section{Observations and results}

\subsection{SAX J1747.0$-$2853}

SAX J1747.0$-$2853 is a recurrent X-ray transient neutron star system first detected by \citet{zand98} and situated only $\simeq$0.3~deg away from Sgr~A. \citet{werner04} studied 45 type-I (thermonuclear) X-ray bursts from this source that were observed between 1998 and 2001, and used these to estimate a distance to the source of 7.5$\pm$1.3~kpc. NIR extinction for this highly obscured region at the adopted distance is estimated to be $A_K = 2.4$ \citep{marshall06}, and
$A_K = 2.34$ through the whole Galaxy \citep{dutra03}, but we caution that the extinction estimates, derived by these authors, were based on the assumption that the reddening law has a shape described by \cite{cardelli89}, which is known to be inaccurate for some regions close to the Galactic Centre \citep{udalski03,geminale04}.

Whenever SAX J1747.0$-$2853 was observed by sensitive X-ray instruments, it was seen to be above quiescence, with the faintest 2--10~keV flux of $1.9\times10^{-11}$~erg~cm$^{-2}$~s$^{-1}$ observed by {\it Chandra} \citep{wijnands02}.  This value is in agreement with more recent {\it Swift} observations \citep{campana09}, and corresponds to a luminosity of $2 \times 10^{35}$~erg~s$^{-1}$, which we accept as a persistent value. \citet{werner04} note, that the binary must be close to the critical luminosity below which the accretion on to the compact object changes from persistent to the transient nature.

\begin{figure*}
\includegraphics[width=0.35\textwidth]{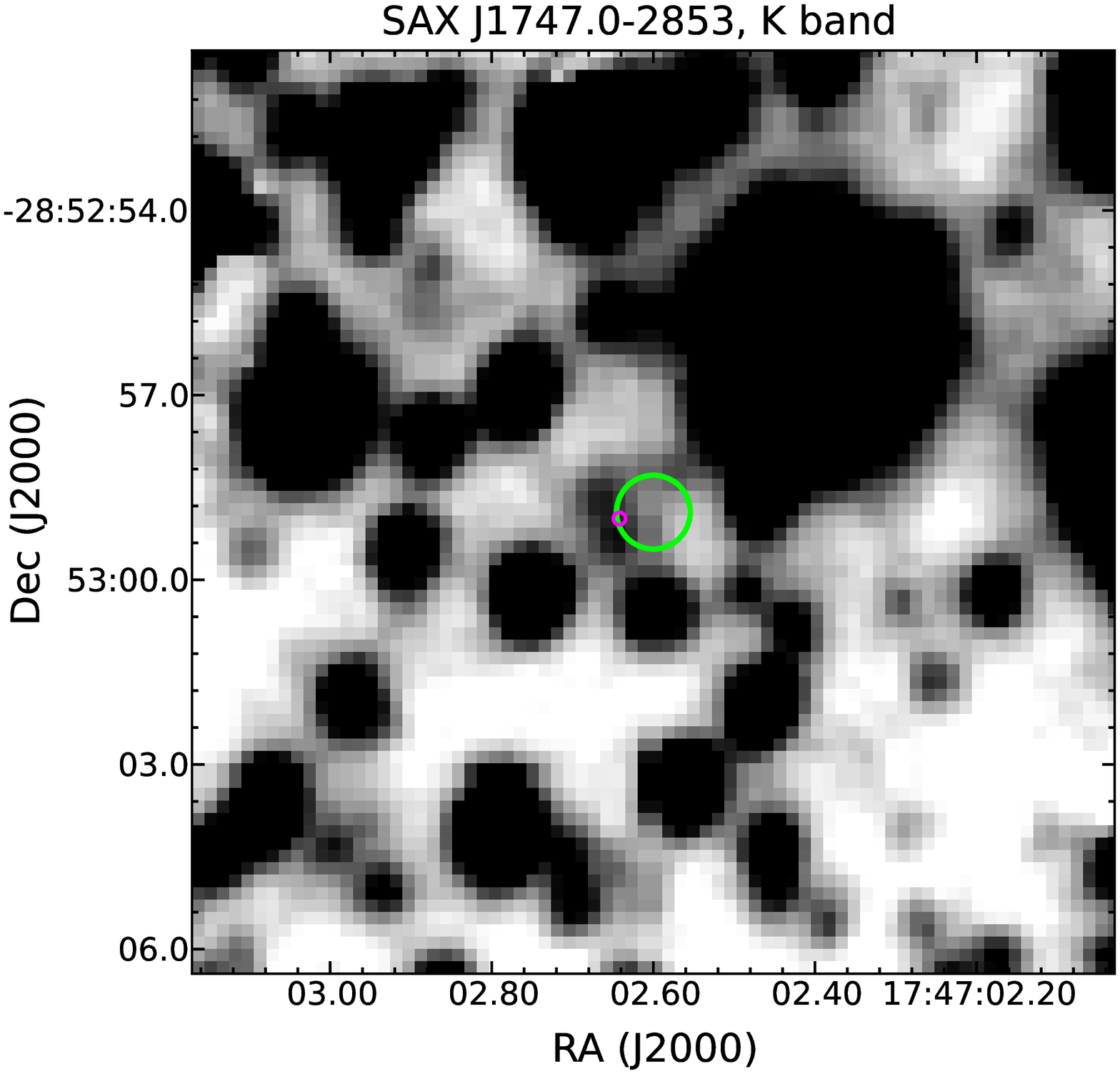}
\includegraphics[width=0.35\textwidth]{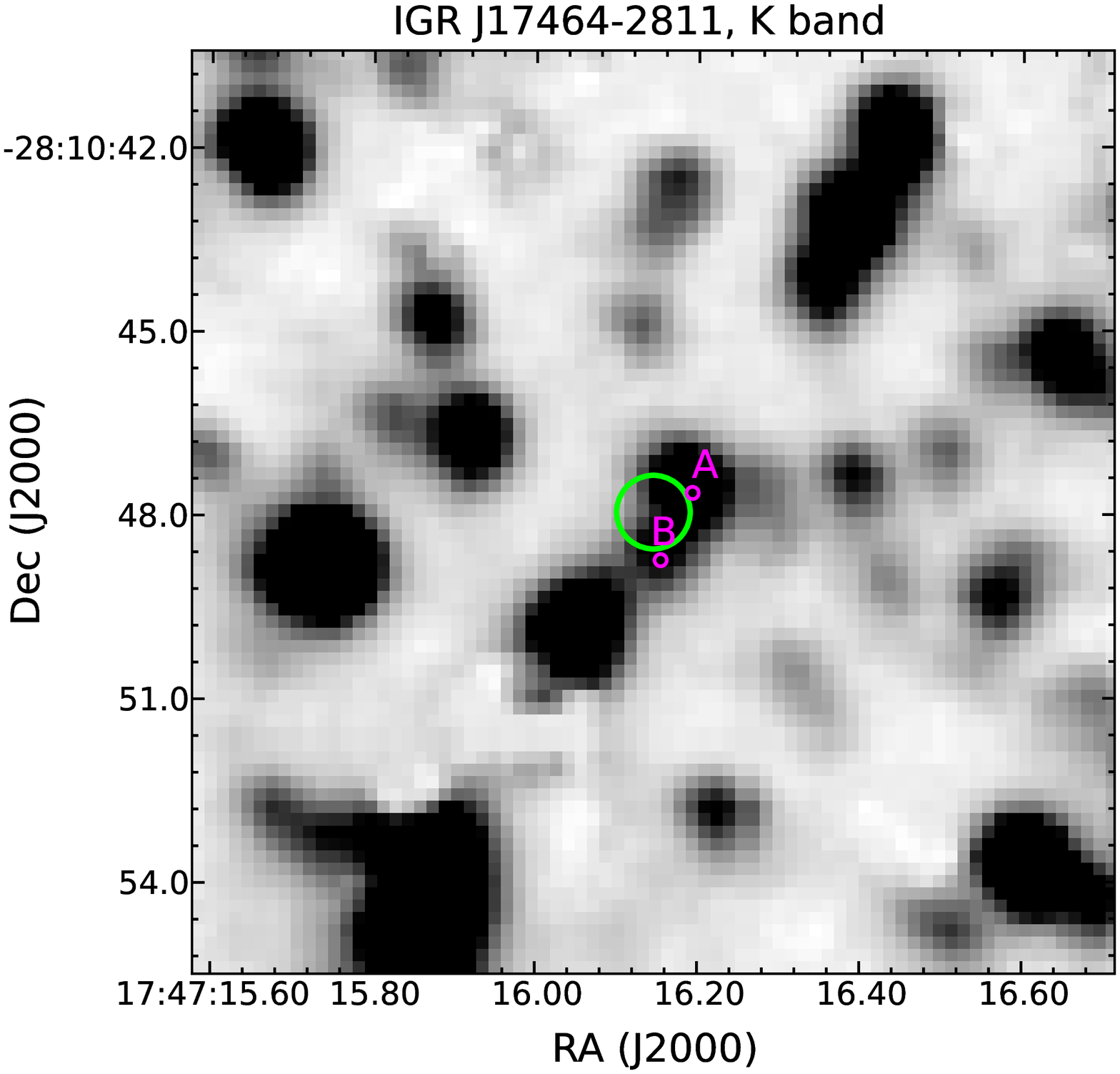}\\
\includegraphics[width=0.35\textwidth]{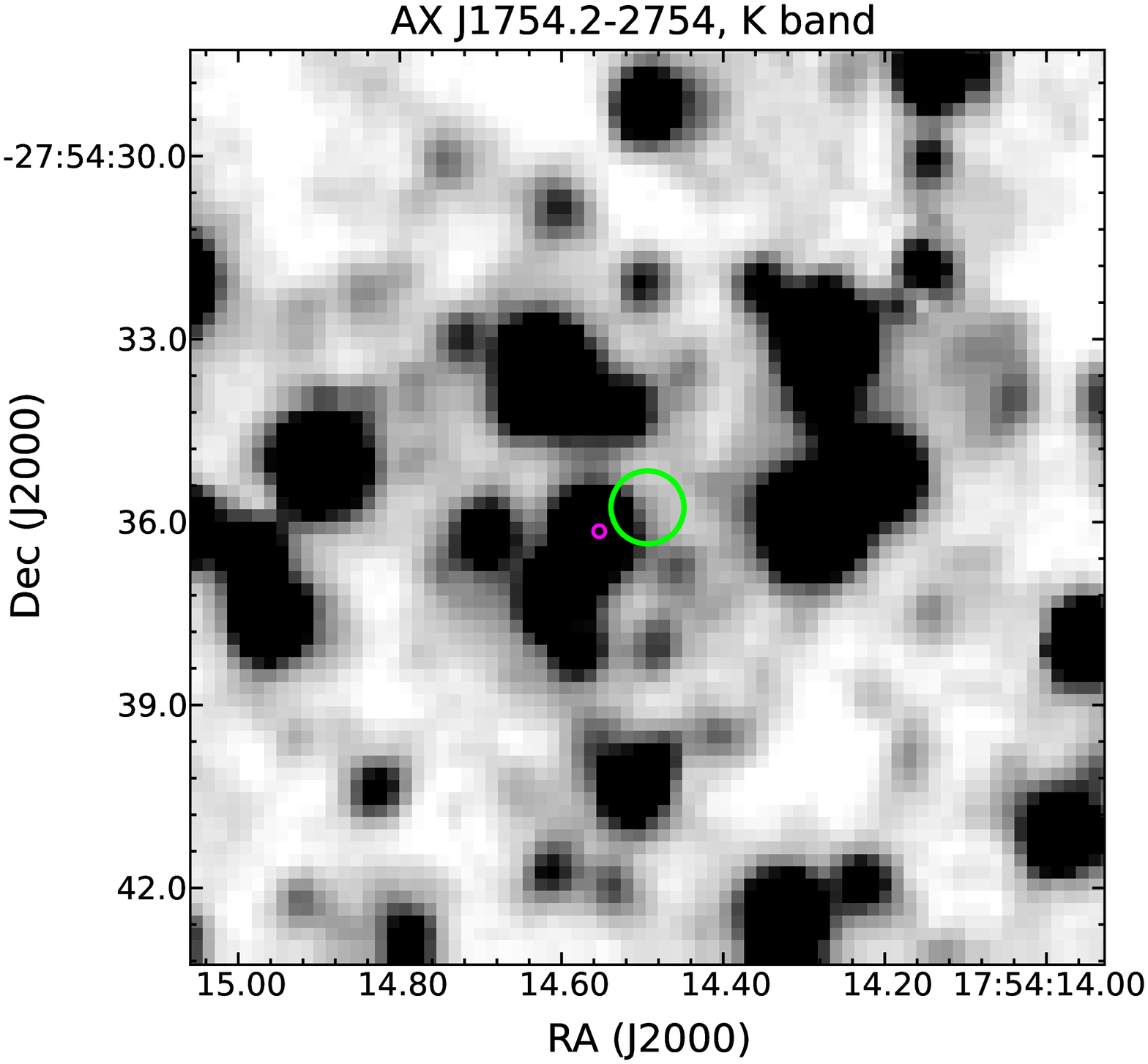}
\includegraphics[width=0.35\textwidth]{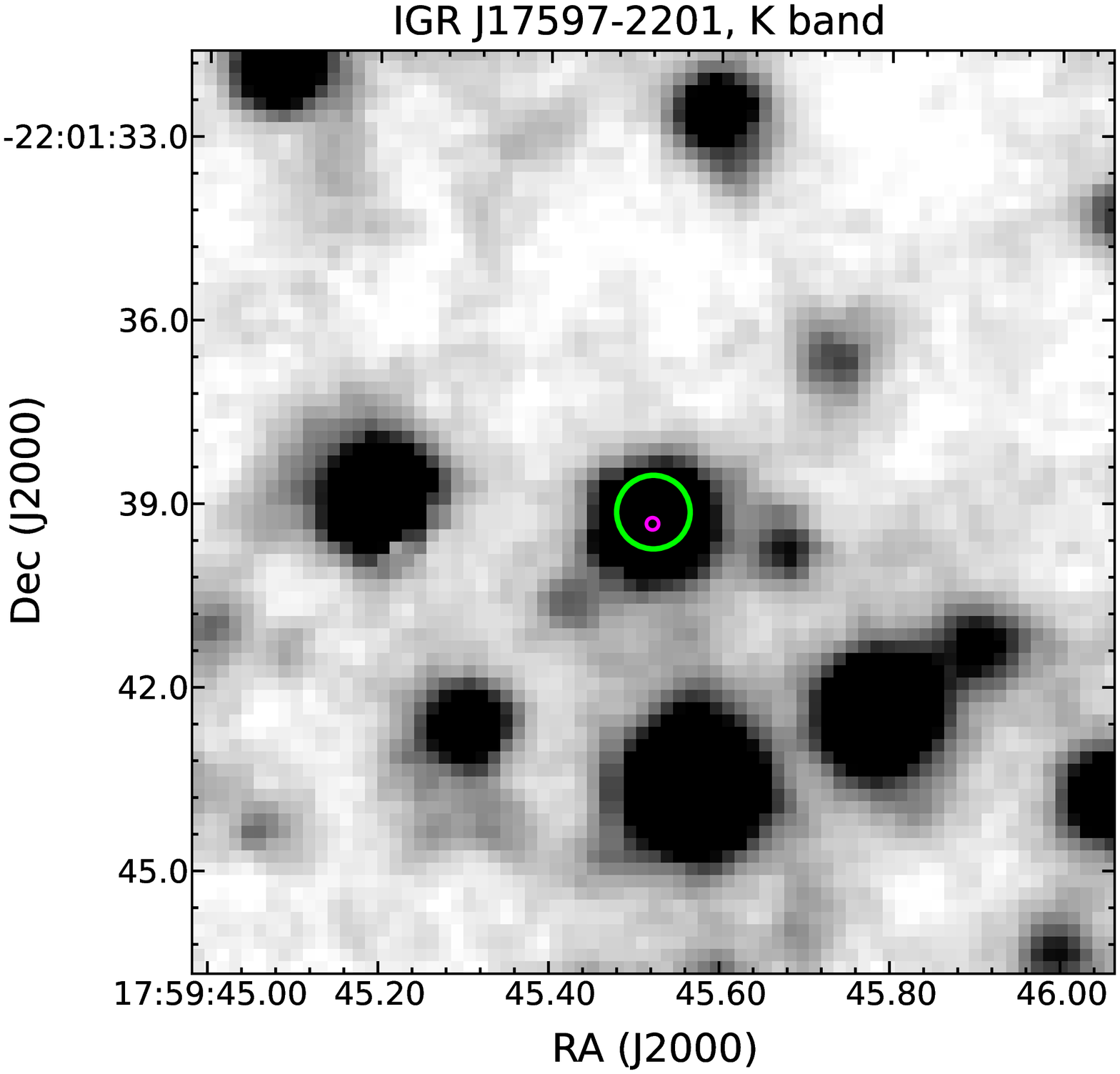}\\
\includegraphics[width=0.35\textwidth]{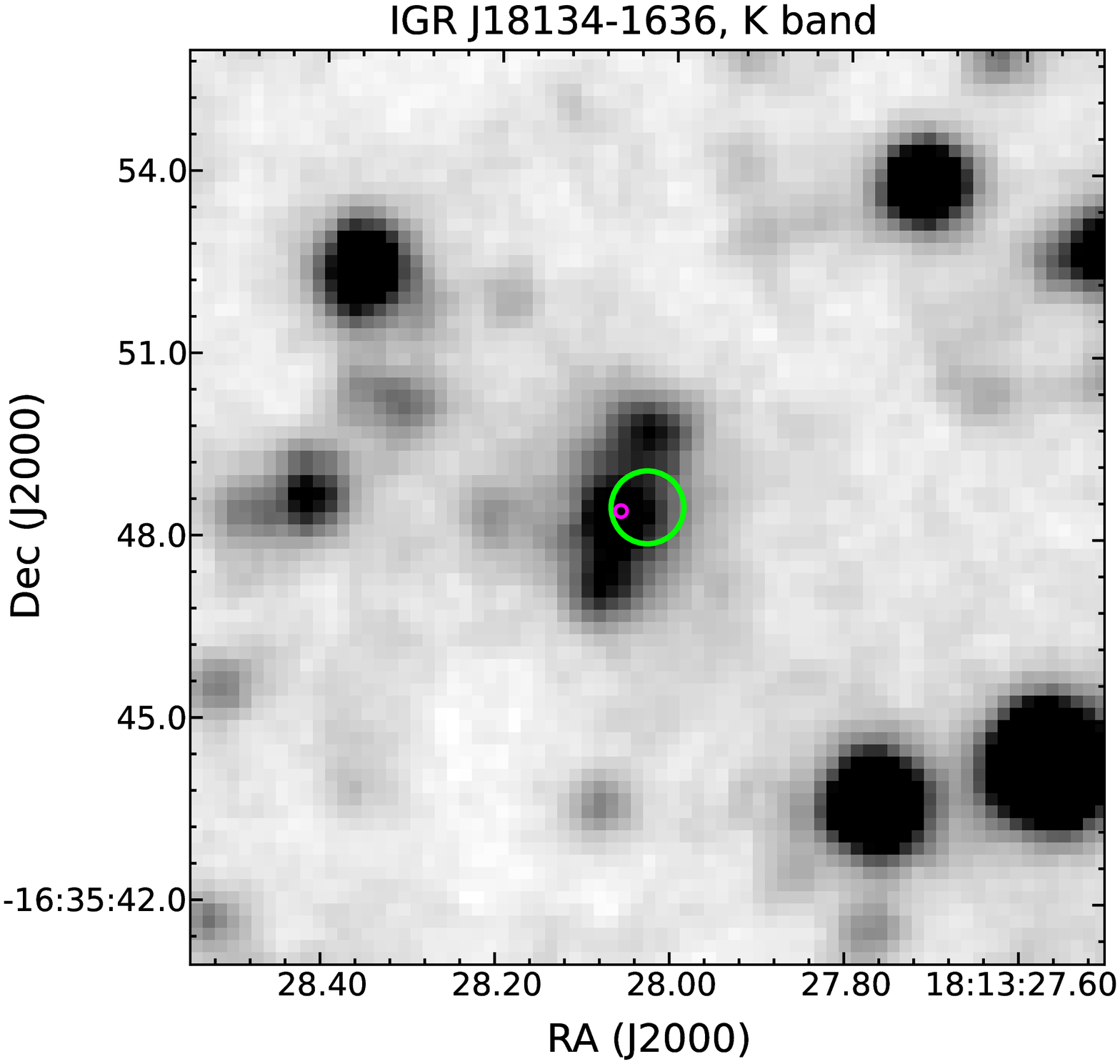}
\includegraphics[width=0.35\textwidth]{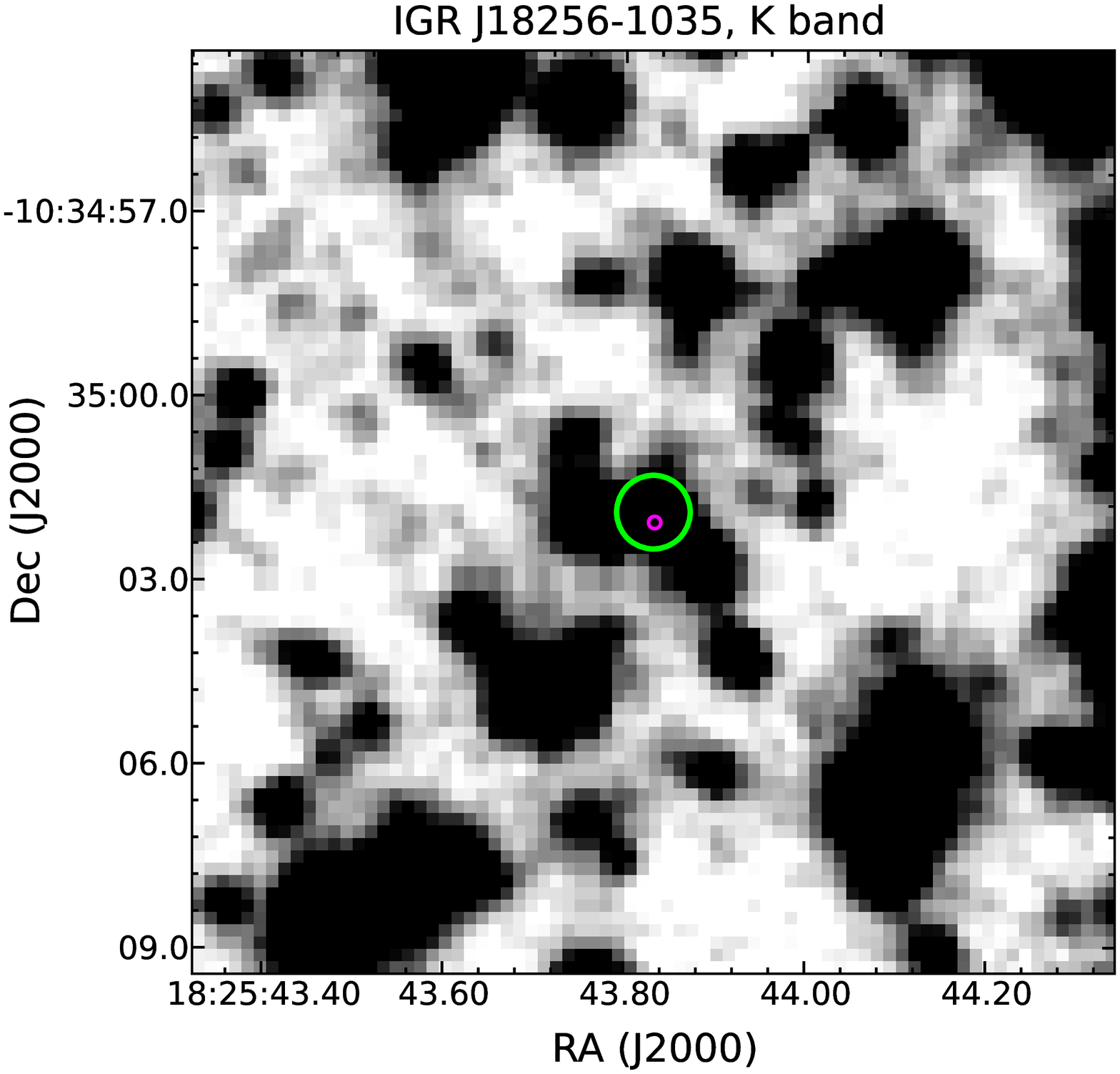}\\
\includegraphics[width=0.35\textwidth]{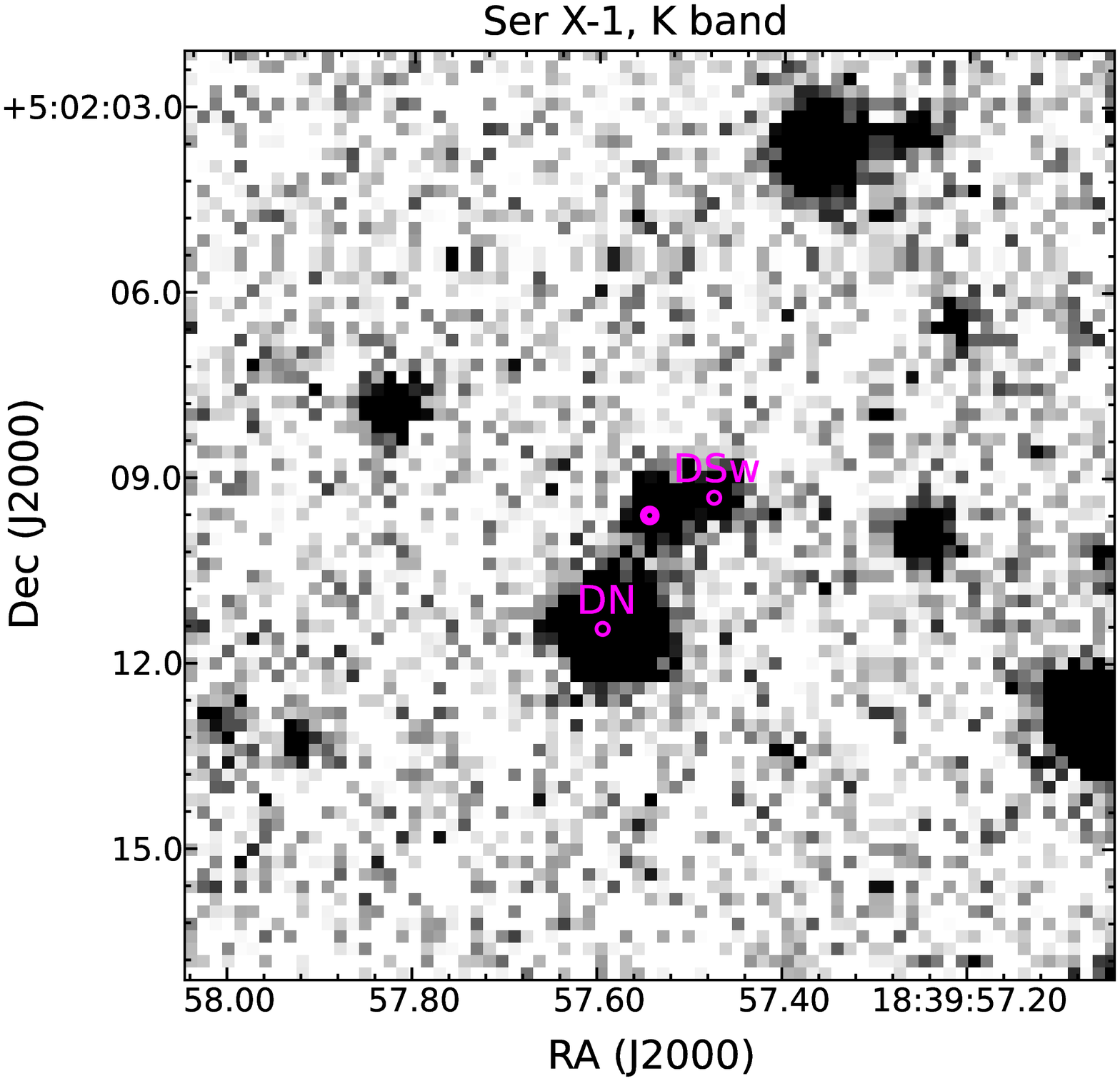}
\caption{NIR images of the studied X-ray sources fields from UKIDSS GPS DR7 with {\it Chandra} $1\sigma$ positional uncertainties overplotted as larger circles in the center. Possible counterparts are marked with smaller magenta circles and discussed in the text. In the case of Ser~X$-$1 its optical counterpart was known before. Solid point in the Ser~X$-$1 field represents known optical counterpart position (usually denoted as DSe), whereas empty points with labels denote nearby objects DN and DSw as per \citet{wachter97}.}
 \label{fig_field}
\end{figure*}

\begin{figure*}
\begin{center}
\includegraphics[width=0.18\textwidth]{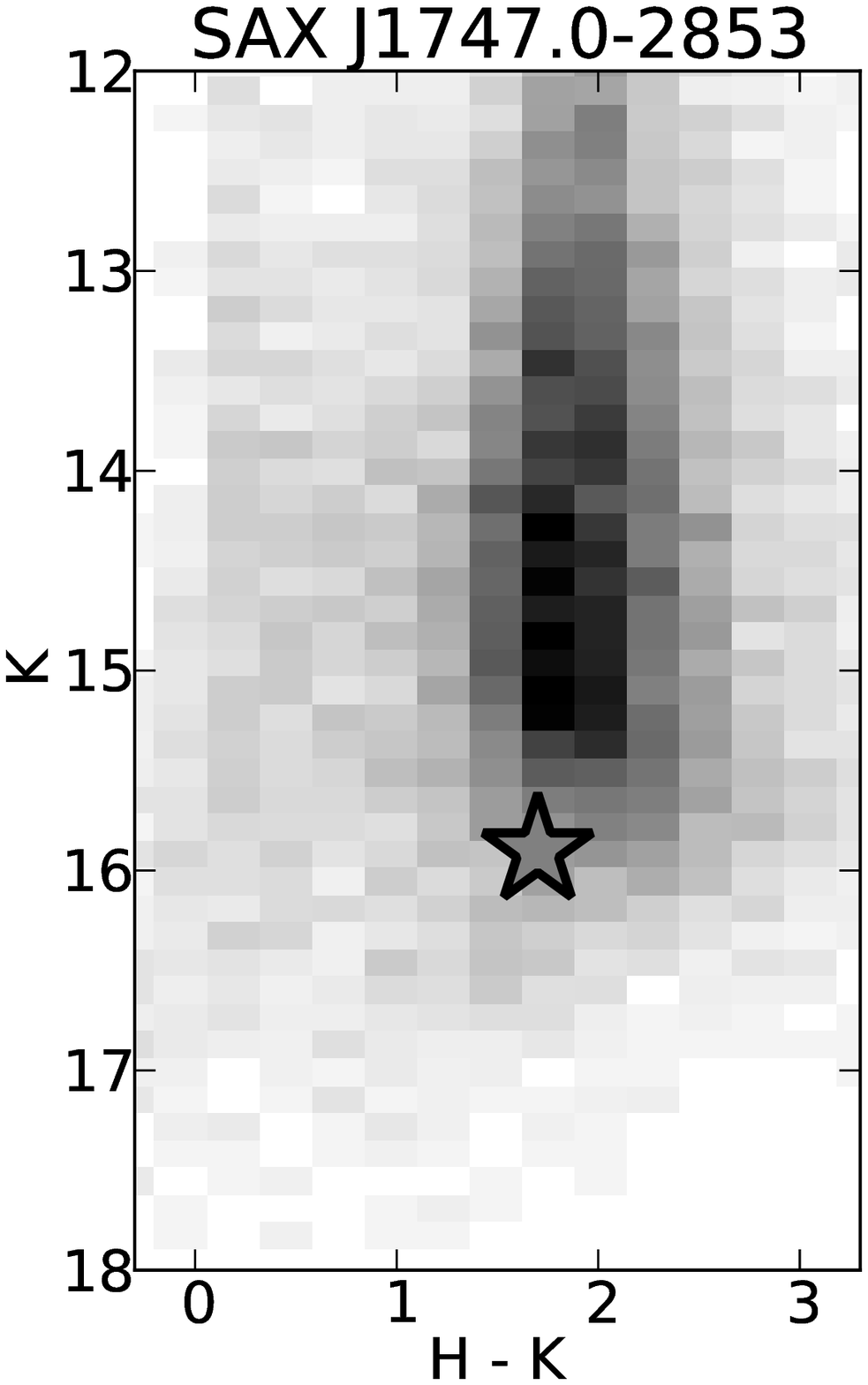}
\includegraphics[width=0.18\textwidth]{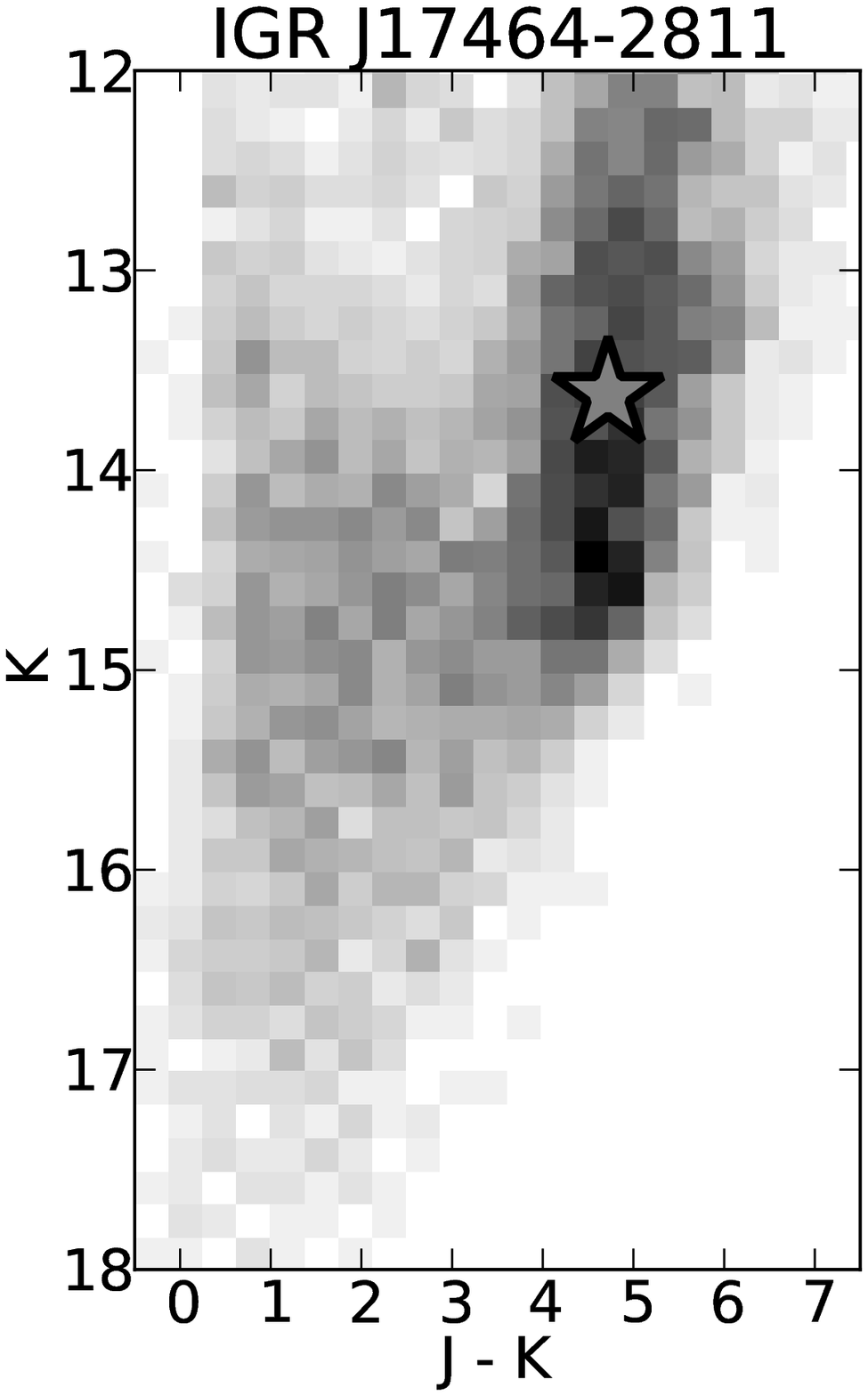}
\includegraphics[width=0.18\textwidth]{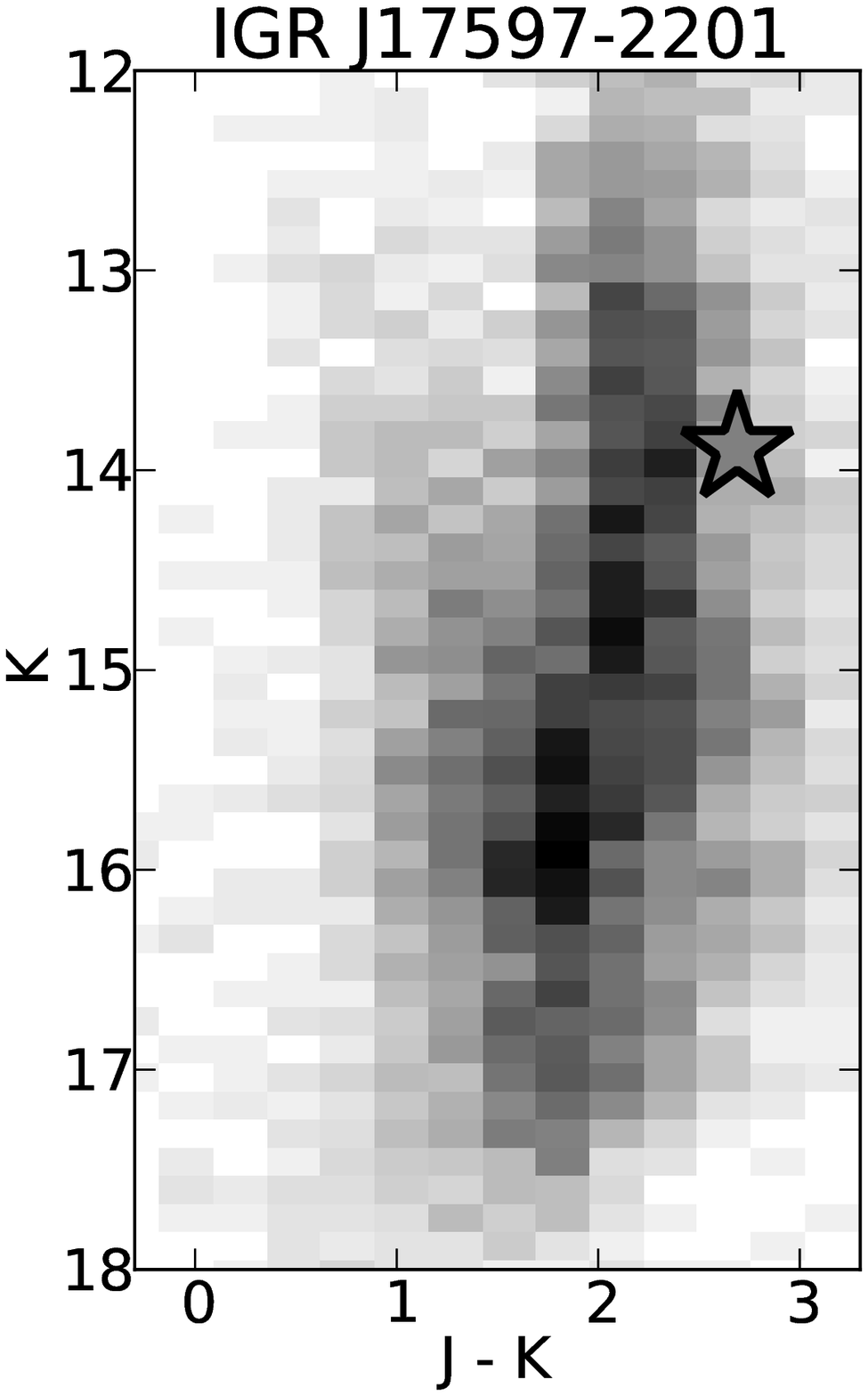}
\includegraphics[width=0.18\textwidth]{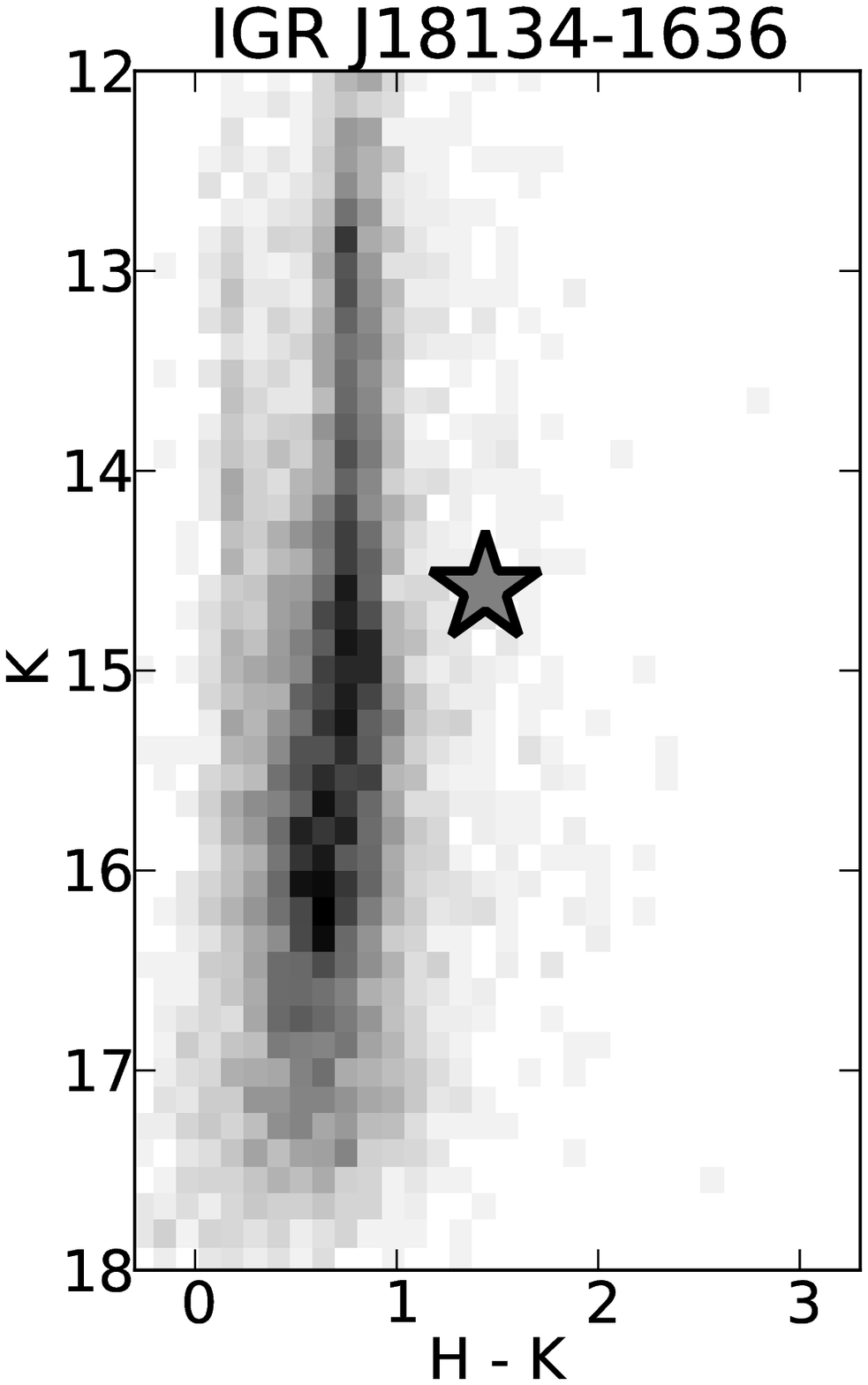}
\includegraphics[width=0.18\textwidth]{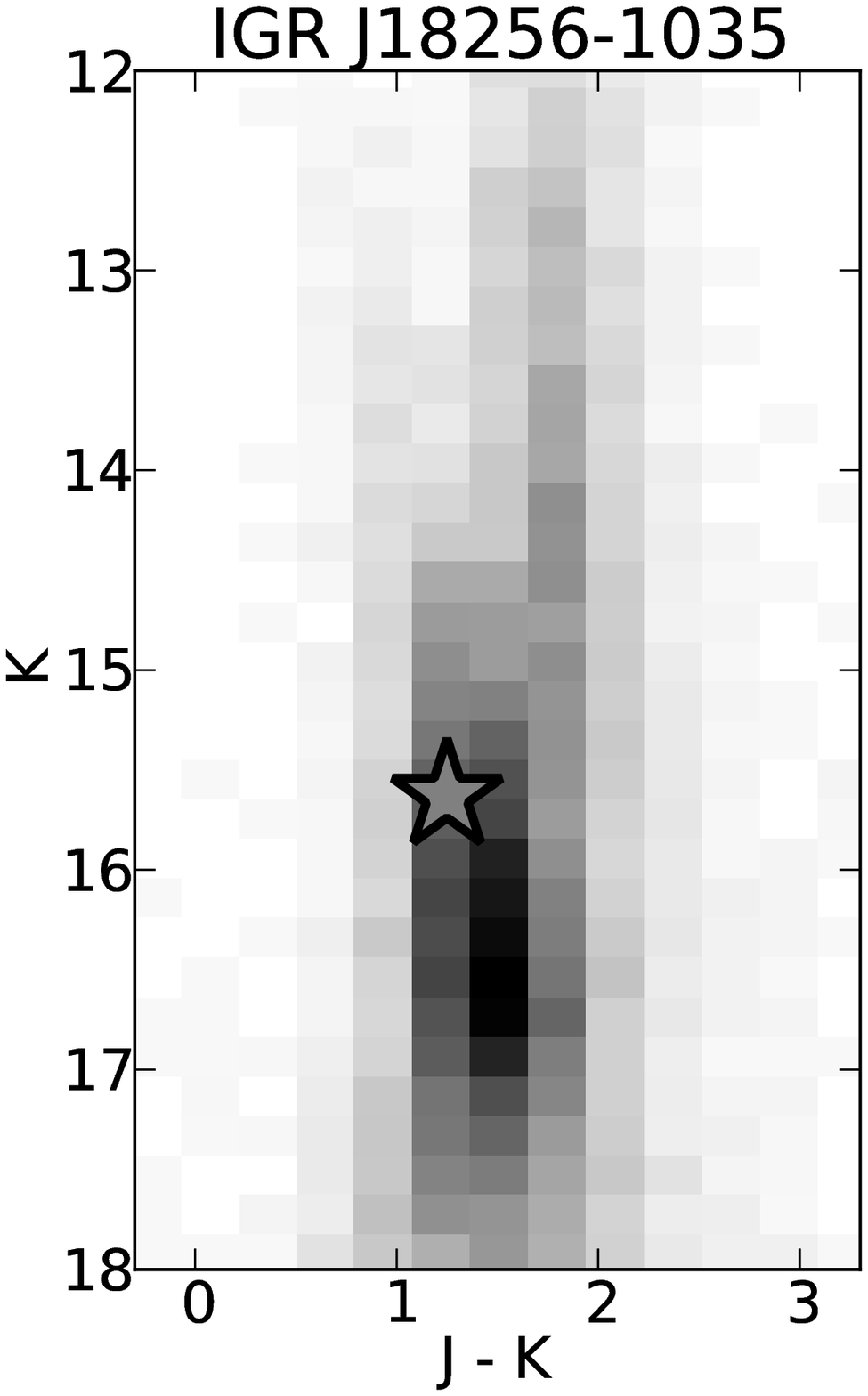}
\end{center}
\caption{NIR colour--magnitude diagrams of 4~arcmin star fields in the vicinities of the studied X-ray sources represented as 2D histograms. Star symbols denote proposed (as in the cases of IGR J17597$-$2201, IGR J18134$-$1636, IGR J18256$-$1035) or unlikely (cases of SAX J1747.0$-$2853, IGR J17464$-$2811) counterparts. See the text for details on each source.
}
 \label{fig_cmd}
\end{figure*}

We used UKIDSS Galactic Plane Survey Data Release 7 data taken by 3.8-m UKIRT telescope to study the field of SAX J1747.0$-$2853 (see Table~\ref{tab_obs} for the observation log). We detected a single source 0.56~arcsec from the X-ray coordinates by \citet{wijnands02} within formal $1\sigma$ X-ray positional uncertainty in $H$ and $K$ filters (see Figure~\ref{fig_field} for the image and Table~\ref{tab_pos} for positional information on this detection). However, this object is not visible in $J$ up to its limiting magnitude. We do not consider this source to be the counterpart of SAX~J1747.0$-$2853 because of the significant coordinates mismatch and because its observed $H-K$ colour is similar to the majority of background stars in this field (see the leftmost panel in Figure~\ref{fig_cmd}). 
Due to different nature of NIR emission we expect the true counterpart to exhibit intrinsically different SED properties with field stars.
As X-ray data show that SAX J1747.0$-$2853 resides inside the Galactic bulge from where most of stars in the CMD come, its true NIR counterpart must experience similar interstellar reddening which cannot significantly influence intrinsic colour difference and hence the observed colours must be different.
Therefore we think that the detected object is a chance projection of a bulge star on to our X-ray error circle.

\begin{table*}
\centering
\caption{X-ray and NIR positional information on the sources analysed in this study. 
For each source (except for Ser X$-$1 for which its optical counterpart was known before) we give X-ray coordinates, their $1\sigma$ uncertainties, and the literature reference for the X-ray position.
Probabilities of a chance background star superposition are given with respect to $K$ filter image and computed with the formula discussed in Section~\ref{sec_data}.
Designation column contains names of possible counterparts how they are discussed in the text. Star symbol denotes suggested counterpart, other variants mean we were not able to suggest a counterpart.
Positional errors determined for possible NIR counterparts in this work are radii of $1\sigma$ coordinate uncertainties in arcseconds with respect to ICRS. 
Last column contains separation between NIR and X-ray positions.}
\label{tab_pos}
\begin{tabular}{lcccccccc}
\thickhline
Source & X-ray position & Pos. err. & Ref.\tnote{a} & Prob. & Desig. & NIR position & Pos. err. & Sep. \\
\hline
 & J2000 & (arcsec) & & & & J2000 &  (arcsec) & (arcsec) \\
\thickhline
SAX J1747.0$-$2853 & 17:47:02.60 $-$28:52:58.9 & 0.6 & [1] & 0.21 & - & 17:47:02.642 $-$28:52:58.97 & 0.2 & 0.56 \\
IGR J17464$-$2811 & 17:47:16.15 $-$28:10:48.0 & 0.6 & [2] & 0.18 & Src A & 17:47:16.195 $-$28:10:47.62 & 0.2 & 0.75 \\
 &  &  & & & Src B & 17:47:16.158 $-$28:10:48.75 & 0.2 & 0.81 \\
AX J1754.2$-$2754 & 17:54:14.49 $-$27:54:35.8 & 0.6 & [2] & 0.20 & - & 17:54:14.553 $-$27:54:36.15 & 0.2 & 0.89 \\
IGR J17597$-$2201 & 17:59:45.52 $-$22:01:39.2 & 0.6 & [3] & 0.11 & * & 17:59:45.522 $-$22:01:39.31 & 0.2 & 0.18 \\
IGR J18134$-$1636 & 18:13:28.03 $-$16:35:48.5 & 0.6 & [4] & 0.12 & * & 18:13:28.059 $-$16:35:48.39 & 0.15 &  0.42 \\
IGR J18256$-$1035 & 18:25:43.83 $-$10:35:01.9 & 0.6 & [5] & 0.19 & * & 18:25:43.836 $-$10:35:02.09 & 0.2 & 0.20 \\
Ser X$-$1                & - & - & - & - & * & 18:39:57.543 +05:02:09.61 & 0.2 & - \\
\thickhline
\end{tabular}
\begin{tablenotes}
\item 1 From \citet{wijnands02}
\item 2 This study
\item 3 From \citet{ratti10}
\item 4 From \citet{tomsick09}
\item 5 From \citet{tomsick08}
\end{tablenotes}

\end{table*}

We also inspected this field in the {\it Spitzer} data archive. 
The nearest object from the GLIMPSE catalogues is 2~arcsec away from the NIR counterpart position and hence is unlikely to be associated with SAX J1747.0$-$2853.
{\it WISE} catalogue does not contain sources within 2~arcsec radius centered at the X-ray position.

SAX J1747.0$-$2853 was reported to exhibit outbursts lasting $\sim$60 days with a recurrence time of 185 days \citep{brandt07} stabilising at approximately the persistent flux level during the intervening period. In particular, it was known to be active in Feb 2006 \citep{chenevez06b} and Sep 2006 \citep{wijnands06b}, so we assume it to be in quasi-quiescence during our NIR observations.

In the absence of counterpart detection we are only able to give the upper limit for SAX J1747.0$-$2853 NIR brightness as the detection limit of used UKIDSS images (see Table~\ref{tab_obs}). This implies lower limit on SAX~J1747.0$-$2853 absolute magnitude of $M_K > 1.0$. In this case, if the secondary fills its Roche lobe, the empirical relation of \cite{revnivtsev12} predicts an upper limit on the orbital period $P \lesssim 50$~hours.

\subsection{IGR J17464$-$2811}
\label{sec_igr_j1746}

Faint accreting neutron star binary IGR J17464$-$2811 (aka XMMU~J174716.1$-$281048) was discovered by {\it XMM-Newton} \citep{sidoli03}. In 2005 {\it INTEGRAL} observed a type-I X-ray burst from this source, which allowed the distance and luminosity of the system to be estimated as $d \simeq 3$~kpc \citet{delsanto07} and $L_{\rm X} = 10^{34}$~erg~s$^{-1}$. According to \citet{marshall06} the line-of-sight extinction for this distance is $A_K \simeq 0.5$~mag. X-ray spectra from this source also demonstrate a high level of interstellar extinction $N_H\simeq6-9\times10^{22}$ cm$^{-2}$ \citep{delsanto07}, which can be translated into $A_V$ and $A_K$  using the relations from \cite{predehl95}, suggesting that $A_K\sim3-5$. Discrepancy between X-ray extinction values and \citet{marshall06} values at 3~kpc distance may be explained by invoking local absorption material.

We studied the field of IGR J17464$-$2811 using UKIDSS GPS DR7 $JHK$ data (see Table~\ref{tab_obs}). Two objects were detected in the vicinity of the X-ray error circle, source A (0.7~arcsec away from its centre) and source B (0.8~arcsec away), both of which have a similar colour $H-K \simeq 1.6$. No sources are visible inside the $1\sigma$ X-ray error circle after subtracting the modelled PSF of sources A and B, so we cannot claim a positional identification. See Figure~\ref{fig_field} for the $K$ band field image and Table~\ref{tab_pos} for positional information on these two detections.

We analysed the brightest and closest visible object, source A, in order to test its association with the X-ray source. 
The {\it Spitzer} archive contains a single source, G000.8344+00.0834, which coincides with source A within coordinate uncertainties, so we consider it to be the same object. 
Source B has no counterpart in the GLIMPSE data.
{\it WISE} catalogue does not contain sources within 2~arcsec radius from the X-ray position.
Hence we were able to construct a 5-band SED of source A and fit it with stellar atmosphere models of \citet{castelli04} with extinction, obtaining best-fitting values of free parameters $T_{\rm eff} = 3800 \pm 200$~K and $A_K = 2.6 \pm 0.1$, reduced $\chi^2 = 0.12$.

We also tried to fit extincted Rayleigh-Jeans law into observed SED of source A: $$F_\nu = \frac{R^2}{D^2} \times \frac{2\pi \nu^2}{c^2} kT \times 10^{-0.4A_\nu}$$ where $F_\nu$ is the observed flux density in erg~cm$^{-2}$~s$^{-1}$~Hz$^{-1}$, $\nu$ the frequency in Hz, $c$ the speed of light, $k$ the Boltzmann constant, $T$ the effective temperature of blackbody emitter in K, $R$ its radius in cm, $D$ the distance in cm, $A\nu$ the line of sight extinction at given frequency, defined by \citet{cardelli89} law with a single parameter $A_V$, the extinction in the optical $V$ band. By fitting this function to the observed SED we obtain 2 parameters, normalization factor comprising combination of $T$, $R$, $D$, and line of sight extinction $A_V$. The latter for convenience we express as the extinction in $K$ band $A_K$, given that $A_K/A_V = 0.11$ \citep{cardelli89}.

\begin{figure}
\includegraphics[width=0.5\textwidth]{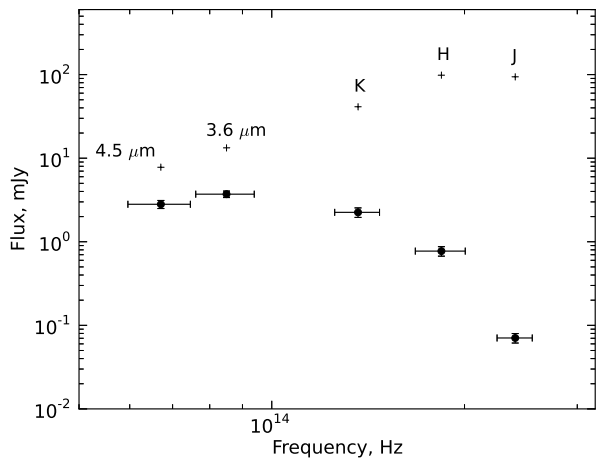}
\caption{Flux versus frequency plot of source A from the vicinity of IGR J17464$-$2811 X-ray error circle. Upper set of points is flux corrected for best-fitting value of Galactic extinction $A_K = 2.6$, lower set of points is observed flux not corrected for extinction. GLIMPSE and UKIDSS data are shown. Photometric errors and bandwidths for compactness are displayed for observed flux values only, but correspondingly apply to the extinction-corrected data points too. Similarly, band labels apply to both sets of points.}
 \label{IGR_J17464-2811_SED}
\end{figure}

The Rayleigh-Jeans law describes the SED of source A with worse statistics than stellar atmosphere models, namely reduced $\chi^2 = 3.2$.
Both extinction-corrected and observed fluxes vs. frequency of source A are presented on Figure~\ref{IGR_J17464-2811_SED}.

Whereas such extinction is intuitively very large, in the \citet{marshall06} 3D extinction catalogue this region is shown to be one of the most obscured in the Galaxy, reaching our inferred values at distance of only 8~kpc. Large reddening values in this direction are also immediately clear from the colour-magnitude diagram of stars in the vicinity of X-ray coordinates (see Figure~\ref{fig_cmd}), because red giant branch on CMD is shifted redwards to $J - K \simeq 4-5$ (note that source A resides within this branch).

On the basis of observed SED of the source A being well represented by a simple cold stellar atmosphere, and its position on the CMD consistent with the red giant branch, we conclude that source A is a background red giant, not associated with IGR J17464$-$2811. Source B is also unlikely a counterpart of IGR J17464$-$2811 because its $JHK$ colours are very similar to source A and hence it is likely to be a background object as well. Therefore, we are unable to associate any existing UKIDSS or GLIMPSE source with IGR J17464$-$2811 and only provide upper limit of its NIR magnitude derived from the detection limit of our $JHK$ images (see Table~\ref{tab_obs}). Deeper observations are required to identify the source in NIR.

Adopting $A_K=0.5$ and the source's distance 3~kpc from \citet{delsanto07}, from the detection limit of the image in $K$ band, $m_{\rm lim} = 17.5$ we impose lower limit on NIR absolute magnitude of the binary system $M_K \gtrsim 4.6$. Then subsituting estimated persistent X-ray luminosity value of $L_{\rm X}=10^{34}$~erg~s$^{-1}$ in period--magnitude relation for persistent LMXBs \citep{revnivtsev12} we give upper limit on the orbital period of IGR J17464$-$2811 (see Table~\ref{tab_summary}).

\subsection{AX J1754.2$-$2754}
The source AX J1754.2$-$2754 was discovered in survey of the Galactic Centre (GC) region by {\it ASCA} observatory \citep{sakano02}. Long term observations of the GC region with {\it INTEGRAL} observatory revealed the source at approximately the same flux level \citep{krivonos07}. The average sources flux is about $10^{-11}$~erg~s$^{-1}$~cm$^{-2}$. In 2005 the type-I X-ray burst was detected from the source, which allowed \citet{chelovekov07} to identify it as a neutron star. The distance to the source was estimated to be $\simeq$7--10 kpc (for different atmosphere models).

The source was observed with {\it Chandra} on July 15, 2008, which allowed us to determine its astrometric position (see Table~\ref{tab_pos}). Additional refinement of the astrometry is not possible due to lack of field stars visible both on X-ray and NIR/optical images.

We examined the field of AX J1754.2$-$2754 in UKIDSS GPS DR7 data. The $JHK$ images contain no positional counterpart within $1\sigma$ X-ray error circle (see Figure~\ref{fig_field}), the nearest object is 0.89~arcsec away (see Table~\ref{tab_pos}). 
This allows us to put only an upper limit on brightness of the NIR counterpart $M_K>2.8$.
We would like to note that the observing night was of a moderate quality so the actual limiting magnitude is worse than typical for UKIDSS datasets (see Table~\ref{tab_obs}).
The only nearby source in the GLIMPSE catalog is 1.9~arcsec away from the X-ray coordinates so is unlikely to be associated with AX J1754.2$-$2754 given {\it Spitzer} positional uncertainty of 0.3~arcsec on each coordinate.
{\it WISE} catalogue does not contain sources within 2~arcsec radius centered at the X-ray position.

Assuming that the secondary star fills its Roche lobe then the upper limit on the NIR brightness can be used to constrain the orbital period of AX J1754.2$-$2754.
Adopting $L_{\rm X}\simeq10^{35}$~erg~s$^{-1}$ and the appropriate scaling relation from \cite{revnivtsev12} we get $P<9$ hours. In fact the persistent nature of the source allows us to put tighter constraints on the orbital period of the system. Adopting the boundary of the persistent luminosity from \cite{dubus01,revnivtsev11} $L_{\rm X}/10^{37}\simeq0.025R_h^{1.4}$ we can get $P\lesssim 0.5$ hours. Allowing the presence of some systematic uncertainties of the adopted transient--persistent--line boundary we can conservatively assume that $P\lesssim 2$~hours.

\subsection{IGR J17597$-$2201}
\label{sec_igr_j1759}

IGR J17597$-$2201 (also sometimes referred as XTE J1759$-$220) is a LMXB hosting a neutron star as suggested by the presence of X-ray bursts \citep{markwardt03} and high/soft spectral state spectrum typical for NS binaries \citep{lutovinov05}. \citet{markwardt03} reported $\simeq30$~per~cent dips with a typical duration of $\simeq5$~minutes which authors interpret as an evidence for 1--3~h orbital period, however it is highly unconstrained. This X-ray binary demonstrates long term variations of its flux (see e.g. its light curve on the web\footnote{\url{http://asd.gsfc.nasa.gov/Craig.Markwardt/galscan/html/XTE_J1759-220.html} or \url{http://hea.iki.rssi.ru/integral/survey/source.php?srcid=086}}), which makes the transient/persistent classification of the source unclear.

\citet{chaty08} tried to look for NIR counterpart of IGR J17597$-$2201 using 4~arcsec {\it XMM-Newton} error circle available at that time and identified 6 possible candidates in their $JHK_s$ data obtained with ESO NTT telescope. \citet{ratti10} later observed much smaller {\it Chandra} error circle in $I$ filter and suggested Candidate 1 of \citet{chaty08} to be the actual positional counterpart, however, without giving reliable photometry due to bad weather conditions during observations.

We inspected UKIDSS GPS DR7 images in the field of IGR J17597$-$2201 (see the observation log in Table~\ref{tab_obs}). There is single bright source visible inside {\it Chandra} error circle (see Figure~\ref{fig_field}) consistent with the position of Candidate 1 from \citet{chaty08} as well as with their photometry. We consider it to be the positional counterpart of IGR J17597$-$2201 and give its photometric parameters in Table~\ref{tab_summary}. We note that as expected the counterpart is redder than most of the stars in its vicinity (see the third panel from the left in Figure~\ref{fig_cmd}).

GLIMPSE catalogue contains single source G007.5695+00.7703 detected at 3.6 and 4.5~microns within $1\sigma$ X-ray positional uncertainty, which we consider to be the same object. Though the catalogue does not include the source at longer wavelengths, it is clearly visible in 5.8~microns image, so we were able to perform PSF photometry and obtain source's magnitude at 5.8~microns and upper limit at 8.0~microns (see Table~\ref{tab_summary}).
{\it WISE} catalogue does not contain sources within 2~arcsec radius centered at the X-ray position.

We fitted extincted Rayleigh-Jeans law into observed SED of source as described in Section~\ref{sec_igr_j1746} in order to test if the observed data is compatible with a typical stellar SED approximation.
For this source we obtained the value of line of sight extinction $A_K = 1.77 \pm 0.07$, the fit had reduced $\chi^2 = 2.8$.
We consider this fit quality satisfactory given the unknown systematical differences between the two surveys used and multi-epoch nature of the data on potentially variable source.

\begin{figure}
\includegraphics[width=0.5\textwidth]{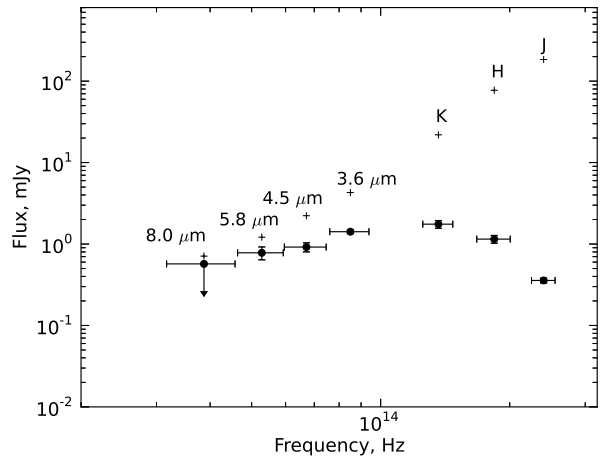}
\caption{Flux versus frequency plot for IGR J17597-2201 counterpart in the near-infrared. Upper set of points is flux corrected for Galactic extinction such that flux distribution matches best Rayleigh-Jeans law, lower set of points is observed flux uncorrected for extinction. GLIMPSE and UKIDSS data are shown. Photometric errors and bandwidths for compactness are displayed for observed flux values only, but correspondingly apply to the extinction-corrected data points too. Similarly, band labels apply to both sets of points.}
 \label{IGR_J17597-2201_SED}
\end{figure}

Overall extinction-corrected and observed fluxes of IGR J17597$-$2201 are presented on Figure~\ref{IGR_J17597-2201_SED}. In this study we adopt the value of extinction $A_K \simeq 2.65$~mag calculated from X-ray data by \cite{chaty08}, though we note that our best-fitting value is slightly smaller.

The distance to this binary system was estimated by \cite{lutovinov05} to be around 5--10 kpc based on the assumption that the luminosity at which the neutron star accreting binaries typically change their spectral state/spectral hardness is universal. 
Adopting this distance one can estimate the absolute brightness of the binary to be $M_K \simeq -2.2--3.1$. 
Such bright NIR counterparts are typical for long period LMXBs with high X-ray luminosities \cite[see e.g.][]{revnivtsev11,revnivtsev12}. 
Assuming the X-ray luminosity of the binary during the analysed NIR observations was $\simeq5\times10^{35}$~erg~s$^{-1}$ and that the secondary star fills its Roche lobe we can estimate the binary system period following approach of \cite{revnivtsev12} to obtain $P\simeq800-2000$ hours. 
Such large orbital period is highly improbable for a system with low persistent X-ray luminosity. 
On the contrary, if the system is transient, then the observed NIR brightness should be attributable to the secondary star. 
In this case the secondary star should be a giant. 
Therefore the only way under assumptions above for such system to produce X-ray emission is an accretion of the stellar wind from giant secondary on to a compact object, a neutron star in the case of IGR~J17597$-$2201.
Systems like this are called symbiotic X-ray binaries (SyXB) and they are known to be rare \citep{masetti07} despite of the theoretical predictions from population synthesis \citep{lu12}.
The wind accretion cannot provide high mass accretion rates and thus limits their X-ray luminosity at the level of $10^{34-35}$~erg~s$^{-1}$.
It is only recently we started to sample with X-ray observations these luminosities at the typical Galactic distance scale of few kiloparsecs, and started to discover new possible SyXBs \citep[e.g.][]{revnivtsev13}.
It is hence natural to expect more SyXB to be discovered in coming years.
We think that IGR~J17597$-$2201 binary system is likely an another example of emerging class of symbiotic X-ray binaries in our Galaxy.
A spectroscopic follow-up observations are required for definitive conclusions on the nature of IGR~J17597$-$2201.

\subsection{IGR J18134$-$1636}

IGR J18134$-$1636 was discovered by {\it INTEGRAL} \citep{bird06,krivonos07}. Its accurate astrometric position was determined from {\it Chandra} observations \citep{tomsick09}. The energy spectrum of the source is highly absorbed, with $N_H \simeq (4-17) \times10^{22}$~cm$^{-2}$.

We studied NIR data for this source in UKIDSS DR7 (see Table~\ref{tab_obs}) and report that single object is visible in $H$ and $K$ band images inside X-ray positional uncertainty, whereas it is not present in $J$ band exposure (see the $K$ band image in Figure~\ref{fig_field}). Based on NIR coordinates from UKIDSS data we also identify this NIR counterpart with the object denoted as G013.8877+00.5968 in the GLIMPSE catalogue where it is detected in all 4 bands. Separation between these two catalogued sources is 0.3~arcsec. Furthermore, source is present in the {\it WISE} \citep{wright10} all-sky catalogue at the nominal distance of 0.6~arcsec from the centre of X-ray coordinates. Since it exhibits similar fluxes as in {\it Spitzer} data and given 0.4~arcsec uncertainty of {\it WISE} position on each coordinate, we consider it to be the same object.

\begin{figure}
\includegraphics[width=0.5\textwidth]{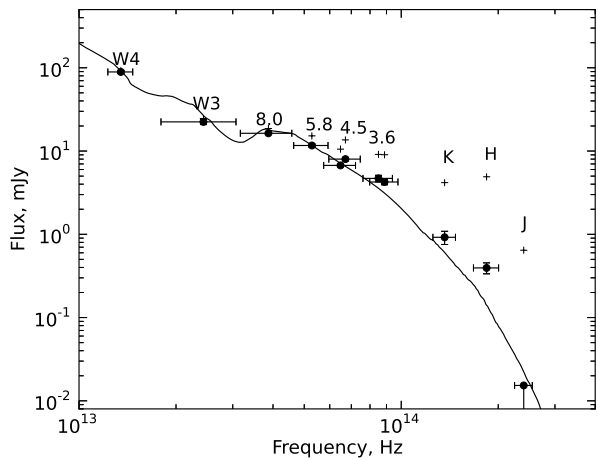}
\caption{Flux versus frequency plot for IGR J18134$-$1636 counterpart in the near-infrared. Upper set of points is flux corrected for best-fitting value of Galactic extinction $A_K = 1.7$, lower set of points is observed flux. Solid line represents a model of heavily obscured broad absorption line quasar Mrk 231 fit to the observed flux. GLIMPSE, UKIDSS and {\it WISE} data are shown.  Photometric errors and bandwidths for compactness are displayed for observed flux values only, but correspondingly apply to the extinction-corrected data points too. Similarly, band labels apply to both sets of points.}
 \label{IGR_J18134-1636_SED}
\end{figure}

We tried to determine source's extinction with the same method we used above for other sources and note that the data cannot be fit with Rayleigh-Jeans law in the physically meaningful range of extinctions. This implies the presence of significant emission excess at MIR wavelengths discrepant from the blackbody radiation, akin to the obscured supergiant X-ray binary IGR J16318$-$4848 \citep{chaty12} which is also known to possess significant local absorption. However, IGR J18134$-$1636 shows 2--3 times larger MIR/NIR flux ratio (see Figure~\ref{IGR_J18134-1636_SED}). We think that this MIR excess feature and distance constraint $d = (L_X/10^{35})^{1/2} \times 17$~kpc make it unlikely for IGR J18134$-$1636 to be an HMXB, but further observations (e.g. infrared spectroscopy) are required to test this idea.

GLIMPSE source G013.8877+00.5968 was also classified as young stellar object (YSO) candidate by \citet{robitaille08} based on its intrinsically red mid-infrared colour. In the NIR the source is indeed redder than most of stars in its vicinity (see the second panel from the right in Figure~\ref{fig_cmd}). 
To check the YSO classification we performed photometry fitting using online YSO SED fitting tool \citep{robitaille07} on the UKIDSS, {\it Spitzer} and {\it WISE} datasets, and managed to achieve a fit with a reduced $\chi^2$ = 10.6 for one of the models from the grid described by a distance $d \simeq 600$~pc and an interstellar extinction $A_K \simeq 3.2$. 
While this model may seem unfavourable based on the fit statistics, we suspect that data errors are somewhat underestimated, because they come from different instruments or, maybe more importantly, from different epochs. 
The potential variablity of the candidate definitely adds scatter to non-simultaneous photometric measurements.
To get an idea of what possible uncertainties are we can compare the fluxes of the candidate between nearby GLIMPSE 3.6~$\mu$m, 4.5~$\mu$m bands and {\it WISE} bands W1 and W2 correspondingly (see Table~\ref{tab_summary}). 
They should match each other within the uncertainties, but in fact we see that there is a difference of a factor 2 to the sum of the formal uncertainties values.
Whatever is the reason for this, e.g. systematic photometric error between the surveys or the variability of the source, such mismatch hints that our $\chi^2$ value maybe overestimated.
So we think that in principle infrared SED of IGR J18134$-$1636 can be described by YSO models, and observed X-ray flux does not contradict the hypothesis of a YSO emitting at $L_X \simeq 10^{32}$~erg~s$^{-1}$, which is not  unreasonable for these sources. 
We however discard hypothesis of a YSO nature of IGR J18134$-$1636 based on its X-ray spectral properties; a hard photon index $\Gamma \simeq 1.4$ as per \citet{tomsick09}, which is atypical for an X-ray emitting YSO, and the requirement of excessive extinction to account for the observed infrared SED. 
Total Galactic extinction in the direction of the source is estimated to be $A_K = 2.4$ \citep{schlegel98}, 2.0 \citep{schlafly11} and 1.9 at 15~kpc \citep{marshall06}, much smaller than obtained best-fitting value.

The hard photon index of IGR J18134$-$1636 hints to its probable AGN nature as these are known to have similar photon indices \citep[see e.g.][]{lin12}.
To test this hypothesis we performed 11 band NIR-to-MIR SED fitting with a set of 25 galaxy and AGN templates from the SWIRE library \citep{polletta07} using foreground extinction and redshift as free parameters. We obtained a best-fitting reduced $\chi^2 = 7.3$ using the heavily obscured broad absorption line quasar Mrk 231 template seen through $A_K = 1.7$~mag Galactic extinction. 
Fit degeneracy does not allow us to constrain its redshift, but it is likely to be less than 0.05. 
Though the fit statistics does not look very convincing, we again note that our formal photometry errors may be underestimated.
This AGN template nevertheless matches our data better than other models we checked (stellar atmospheres, Rayleigh-Jeans, YSOs and all other SWIRE library models).
Moreover, obtained value of foreground extinction is in line with the total Galactic extinction in IGR J18134$-$1636 direction and heavily obscured nature of the best chosen template agrees well with X-ray data showing significant intrinsic absorption of the source. 
Therefore we suggest IGR J18134$-$1636 to be an obscured AGN seen through the disc of our Galaxy.

\subsection{IGR J18256$-$1035}

The source was discovered by {\it INTEGRAL} \citep{bird06,krivonos07}. Its astrometric position was refined later by \cite{landi07} using data of {\it Swift} observatory, and the most accurate position was obtained with the help of {\it Chandra} observations by \cite{tomsick08} who estimated 0.3--10~keV flux to be $2.9 \times 10^{-12}$~erg~cm$^{-2}$~s$^{-1}$.

We performed analysis of IGR J18256$-$1035 field in the UKIDSS GPS DR7 data (see observation log in Table~\ref{tab_obs}). 
Single object is visible in these $JHK$ images inside {\it Chandra} positional uncertainty (see Figure~\ref{fig_field}) and we propose it as the NIR counterpart of this binary. 
The GLIMPSE catalogue contains an object G020.5947+00.8120 detected in single 3.6~$\mu$m band and separated from the centre of X-ray position by only 0.5~arcsec. 
Though formally it may be a counterpart as well, we however think that due to worse angular resolution of {\it Spitzer} IRAC detector compared to the UKIRT, this GLIMPSE object is a blend of 3 sources clearly visible in the $K$ band at the Figure~\ref{fig_field} around X-ray coordinates uncertainty and hence its photometric measurements cannot be used together with the UKIDSS ones. 
Visually elongated shape of the GLIMPSE object supports this idea as well.
Moreover, if we consider 4 band SED ($JHK$ and 3.6~$\mu$m), it cannot be described by a reasonable smooth NIR emission model that fits $JHK$ data as 3.6~$\mu$m band flux is 1.5~mag brighter than simple $JHK$ extrapolation to 3.6~$\mu$m.
{\it WISE} catalogue does not contain sources within 2~arcsec radius centered at the X-ray position.

The source's X-ray spectra are strongly absorbed along the line-of-sight. 
In the work of \citet{tomsick08} the absorption column was determined to be $\simeq3.1\times10^{22}$~cm$^{-2}$ for a power law energy spectrum with photon index $\Gamma=0.1$. 
At the same time the X-ray spectrum presented by \citet{tomsick08} hints for a steeper photon index.
If this is the case the absorbing column value will likely be a bit larger.

The source's $JHK$ SED can be fit with the extincted Rayleigh-Jeans law which results in $A_K = 1.0$ (reduced $\chi^2=1.3$). Using the \citet{marshall06} 3D Galactic extinction model, $A_K = 1.0$ corresponds to the distances larger than 10~kpc for given sky position. The total value of Galactic extinction in this direction in \citet{schlegel98} is smaller than in \citet{marshall06}, $A_K = 1.1$ \citep[compare to $A_K = 0.9$ in][]{schlafly11}. Large photoabsorption and large extinction along the line-of-sight to the source (compared to the total Galactic value) can be taken as an indication that the distance to the source is greater than $\simeq5-10$~kpc. 

At the same time, if we consider the LMXB accretion disc model of \citet{revnivtsev12}, such a combination of X-ray flux, observed $K$ band magnitude and realistic range of extinctions yields orbital periods between 200 and 1000~hours and excessive values of (unreddened) $J-K > 4$ colour within plausible distance. We interpret this as an argument against LMXB nature of this source. For $A_K \simeq 1.0$ and observed colour $J-K=1.25$ the object must be intrinsically blue like e.g. star of a spectral class B having unreddened $(J-K)_0 = -0.25$. 

Based on observed source's colour and reasons about its extinction above, we hence suggest it is HMXB.
Absolute magnitude of a B giant $M_K \simeq -2$ \citep{pickles98} would imply its distance of about 20~kpc and a height of 300 pc above the Galactic plane, which seems excessive for this type of object \citep{lutovinov13}. On the other hand we note that IGR J18256$-$1035 is situated only 50~arcmin away from nearest star-forming complex 20.7-0.1 in \citet{russeil03} catalogue, which may agree with the correlation between HMXBs and the distribution of OB star-forming complexes \citep{bodaghee12,coleiro13}. The kinematic distances to 20.7-0.1 (11.8~kpc) and to the second nearest complex 21.0+0.0 (14.0~kpc) approximately 55~arcmin away may be used as indicative for the IGR J18256$-$1035 if it is fainter than a BIII star.

We stress that without additional data it is not possible to rule out a chance superposition of a nearby B star, inactive in X-rays, that out-shine the (fainter) LMXB.

\subsection{Ser X$-$1}

Ser X$-$1 was discovered using rocket experiments by \cite{bowyer65}. Detection of type-I X-ray bursts from the source established that the binary harbours a neutron star \citep{swank76,li77}. The source distance was estimated to be $9.5-12.7$ kpc \citep{jonker04}.
Its average luminosity is about $5\times10^{36}$~erg~s$^{-1}$ \citep{masetti04}.

It has been long thought that the optical counterpart of Ser X$-$1 is an MM Ser variable. However, this turned out to be a close superposition of several objects. First \citet{thorstensen80} found that the previously identified counterpart contains two components, unrelated north DN and true variable south DS, separated by 2.1~arcsec. Later \citet{wachter97} using PSF fitting showed that the southern component of the blend DS in turn contains two optical objects, eastern DSe and western DSw, separated by around 1~arcsec. \citet{wachter97} were not able to reliably measure DSe photometry, however they proposed it to be the true counterpart of Ser X$-$1 based on its bluer colour. Later \citet{hynes04} carried out high-quality spectral observations of the DSe and DSw blend and confirmed that DSe is the true LMXB based on its spectral features.

The field of Ser X$-$1 was twice observed by UKIDSS (see Table~\ref{tab_obs}), during the first epoch in the $K$ band only (included in DR7 and later) and during the second epoch in all three $JHK$ filters (available starting from DR9 release). Both nights had excellent seeing conditions. We reliably detect and resolve all the close objects previously known to compose MM Ser including true counterpart DSe (see Figure~\ref{fig_field}) and we list its first NIR photometry estimates in Table~\ref{tab_summary}. Two $K$ band magnitudes favor for slight variability of the source though its confidence is not so certain.

We must also note that dereddened colour of Ser X$-$1 $J-K = -0.1$ well matches value $J-K \simeq 0.0$ expected for this system from \citet{revnivtsev12} irradiated accretion disc model in X-ray binaries with Roche-lobe overflow.

GLIMPSE catalogue does not contain any sources within 2~arcsec from the optical position of Ser X$-$1.

\section{Summary}
\label{sec_summary}
We studied data of three infrared surveys, UKIDSS, GLIMPSE and {\it WISE}, around positions of 7 X-ray sources. All them but two were unidentified in optical before and had no certain type identification. In 4 cases we were able to detect their positional counterparts in arcsec-scale X-ray error circles and in 3 cases we determined an upper limit for the source's brightness. Based on this multi-band photometric information we suggest that 1 source from our set is a rare symbiotic X-ray binary, 4 are LMXBs (one of them, Ser X$-$1 is well-studied, but had no NIR photometry before), 1 is a HMXB and 1 is an tentative AGN. For one of LMXBs, AX~J1754.2$-$2754, which was not detected directly, we give 2~h upper limit on its orbital period estimated using \citet{revnivtsev12} relation for NIR luminosity of illuminated accretion disc powered through Roche-lobe overflow. Our main results are summarized in Table~\ref{tab_pos} with the positional information and Table~\ref{tab_summary} which gives details on counterparts photometry, suggested types, absolute magnitudes and orbital periods.

\begin{table*}
\centering
\renewcommand{\tabcolsep}{1mm}
\caption{Summary of the NIR and MIR photometry or upper limits and orbital period constraints of the sources analysed in this study. Magnitudes (either positional candidates or upper limits) are given in the Vega system as it is observed with no extinction correction applied. Magnitude errors are $1\sigma$ uncertainties. Absolute magnitudes $M_K$ are calculated for $K$ filter using extinction $A_K$ from \citet{marshall06} or indicated references and distance estimates from the literature (see text). Orbital periods of LMXBs are estimated in hours using \citet{revnivtsev12} relation and X-ray luminosities necessary for that taken from the references, discussed in the text.}
\label{tab_summary}
\begin{threeparttable}
\begin{tabular}{lccccc}
\thickhline
Source & Type & Observed magnitudes & Extinction $A_K$ & Abs. mag. $M_K$ & Period \\
\hline
 & & (Vega mag) & & (Vega mag) & (h) \\ 
\thickhline

SAX J1747.0$-$2853 & NS LMXB & $J > 20.2; H > 19.0; K > 17.8$ & 2.4\tnote{a} & $\gtrsim 1.0$ & $\lesssim 50$\\

IGR J17464$-$2811 & NS LMXB & $J > 20.0; H > 18.7; K > 17.5$ & 0.5\tnote{a} & $\gtrsim 4.6$ & $\lesssim 4$\\

AX J1754.2$-$2754 & NS LMXB & $J > 18.7; H > 18.7; K > 18.0$ & 0.9\tnote{a} & $\gtrsim 2.2-2.6$ & $\lesssim 2$\\

IGR J17597$-$2201& SyXB & $J = 16.58 \pm 0.07; H = 14.87 \pm 0.11; K = 13.89 \pm 0.11$ & 2.65\tnote{b} & $\simeq -2.2...-3.1$ & - \\
 &  & $m_{3.6}=13.244 \pm 0.054; m_{4.5}=13.230 \pm 0.127$ &  &  & \\
 &  & $m_{5.8}=12.93 \pm 0.18; m_{8.0}>12.6$ &  &  & \\

IGR J18134$-$1636 & AGN & $J > 20.0; H = 16.03 \pm 0.15; K = 14.59 \pm 0.18$ & $\simeq 1.9$\tnote{a} & - & - \\
 &  & $m_{3.6}=11.943 \pm 0.089; m_{4.5}=10.879 \pm 0.048$ &  &  & \\
 &  & $m_{5.8}=9.985 \pm 0.042; m_{8.0}=8.984 \pm 0.024$ &  &  & \\
 &  & $m_{W1}=12.156 \pm 0.074; m_{W2}=11.019 \pm 0.036$ &  &  & \\
 &  & $m_{W3}=7.876 \pm 0.072; m_{W4}=4.931 \pm 0.043$ &  &  & \\

IGR J18256$-$1035 & HMXB & $J = 16.88 \pm 0.15; H = 16.16 \pm 0.16; K = 15.63 \pm 0.15$ & $\simeq 1$\tnote{c} & $\simeq -1...-2$ & - \\

Ser X$-$1 & NS LMXB & $J = 17.34 \pm 0.10$; $H = 17.27 \pm 0.06$ & 0.3\tnote{a} & $\simeq 2.1$ & $\simeq 7.0$\\
 &  & $K_{\rm epoch1} = 17.01 \pm 0.06$ &  &  & \\
 &  & $K_{\rm epoch2} = 17.14 \pm 0.09$ &  &  & \\
\thickhline
\end{tabular}
\begin{tablenotes}
\item [a] From \citet{marshall06}
\item [b] From \citet{chaty08}
\item [c] From Rayleigh-Jeans law fitting, this study
\end{tablenotes}

\end{threeparttable}
\end{table*}

\section*{Acknowledgments} 
This research has made use of the VizieR catalogue access tool, CDS, Strasbourg, France. Authors made use a web-site ``INTEGRAL sources'' of Jerome Rodriguez. This publication makes use of data products from the Wide-field Infrared Survey Explorer, which is a joint project of the University of California, Los Angeles, and the Jet Propulsion Laboratory/California Institute of Technology, funded by the National Aeronautics and Space Administration. This work is based in part on observations made with the Spitzer Space Telescope, which is operated by the Jet Propulsion Laboratory, California Institute of Technology under a contract with NASA. 
Authors are grateful to Mark Burke for suggesting a number of improvements and to anonymous referee whose comments helped to improve significantly the contents and the presentation of the manuscript.
IZ was supported by the Russian Foundation for Basic Research grant 12-02-00186.
MR was supported by the RFBR grant 13-02-00741.

\bibliographystyle{mn2e}
\bibliography{lmxb_2013a_v7}

\begin{thebibliography}{73}
\expandafter\ifx\csname natexlab\endcsname\relax\def\natexlab#1{#1}\fi

\bibitem[{{Benjamin} {et~al.}(2003){Benjamin}, {Churchwell}, {Babler}, {Bania},
  {Clemens}, {Cohen}, {Dickey}, {Indebetouw}, {Jackson}, {Kobulnicky},
  {Lazarian}, {Marston}, {Mathis}, {Meade}, {Seager}, {Stolovy}, {Watson},
  {Whitney}, {Wolff}, \& {Wolfire}}]{benjamin03}
{Benjamin}, R.~A., {et~al.} 2003, \pasp, 115, 953

\bibitem[{{Bird} {et~al.}(2006){Bird}, {Barlow}, {Bassani}, {Bazzano},
  {B{\'e}langer}, {Bodaghee}, {Capitanio}, {Dean}, {Fiocchi}, {Hill}, {Lebrun},
  {Malizia}, {Mas-Hesse}, {Molina}, {Moran}, {Renaud}, {Sguera}, {Shaw},
  {Stephen}, {Terrier}, {Ubertini}, {Walter}, {Willis}, \& {Winkler}}]{bird06}
{Bird}, A.~J., {et~al.} 2006, \apj, 636, 765

\bibitem[{{Bodaghee} {et~al.}(2012){Bodaghee}, {Tomsick}, {Rodriguez}, \&
  {James}}]{bodaghee12}
{Bodaghee}, A., {Tomsick}, J.~A., {Rodriguez}, J., \& {James}, J.~B. 2012,
  \apj, 744, 108

\bibitem[{{Bowyer} {et~al.}(1965){Bowyer}, {Byram}, {Chubb}, \&
  {Friedman}}]{bowyer65}
{Bowyer}, S., {Byram}, E.~T., {Chubb}, T.~A., \& {Friedman}, H. 1965, Science,
  147, 394

\bibitem[{{Brandt} {et~al.}(2007){Brandt}, {Chenevez}, {Kuulkers}, {Natalucci},
  {Fiocchi}, {Tarana}, {Shaw}, {Beckmann}, {Courvoisier}, {Domingo}, {Ebisawa},
  {Jonker}, {Kretschmar}, {Markwardt}, {Oosterbroek}, {Paizis}, {Risquez},
  {Sanchez-Fernandez}, \& {Wijnands}}]{brandt07}
{Brandt}, S., {et~al.} 2007, The Astronomer's Telegram, 1228, 1

\bibitem[{{Campana} {et~al.}(2009){Campana}, {Chenevez}, \&
  {Kuulkers}}]{campana09}
{Campana}, S., {Chenevez}, J., \& {Kuulkers}, E. 2009, The Astronomer's
  Telegram, 1951, 1

\bibitem[{{Campana} {et~al.}(1995){Campana}, {Stella}, {Mereghetti}, \&
  {Colpi}}]{campana95}
{Campana}, S., {Stella}, L., {Mereghetti}, S., \& {Colpi}, M. 1995, \aap, 297,
  385

\bibitem[{{Cardelli} {et~al.}(1989){Cardelli}, {Clayton}, \&
  {Mathis}}]{cardelli89}
{Cardelli}, J.~A., {Clayton}, G.~C., \& {Mathis}, J.~S. 1989, \apj, 345, 245

\bibitem[{{Castelli} \& {Kurucz}(2004)}]{castelli04}
{Castelli}, F. \& {Kurucz}, R.~L. 2004, ArXiv Astrophysics e-prints

\bibitem[{{Chaty} \& {Rahoui}(2012)}]{chaty12}
{Chaty}, S. \& {Rahoui}, F. 2012, \apj, 751, 150

\bibitem[{{Chaty} {et~al.}(2008){Chaty}, {Rahoui}, {Foellmi}, {Tomsick},
  {Rodriguez}, \& {Walter}}]{chaty08}
{Chaty}, S., {Rahoui}, F., {Foellmi}, C., {Tomsick}, J.~A., {Rodriguez}, J., \&
  {Walter}, R. 2008, \aap, 484, 783

\bibitem[{{Chelovekov} \& {Grebenev}(2007)}]{chelovekov07}
{Chelovekov}, I.~V. \& {Grebenev}, S.~A. 2007, Astronomy Letters, 33, 807

\bibitem[{{Chenevez} {et~al.}(2006){Chenevez}, {Shaw}, {Kuulkers}, {Brandt},
  {Courvoisier}, {Ebisawa}, {Kretschmar}, {Markwardt}, {Mowlavi},
  {Oosterbroek}, {Orr}, {Paizis}, {Sanchez-Fernandez}, {Wijnands}, \&
  {Zurita}}]{chenevez06b}
{Chenevez}, J., {et~al.} 2006, The Astronomer's Telegram, 734, 1

\bibitem[{{Churchwell} {et~al.}(2009){Churchwell}, {Babler}, {Meade},
  {Whitney}, {Benjamin}, {Indebetouw}, {Cyganowski}, {Robitaille}, {Povich},
  {Watson}, \& {Bracker}}]{churchwell09}
{Churchwell}, E., {et~al.} 2009, \pasp, 121, 213

\bibitem[{{Coleiro} \& {Chaty}(2013)}]{coleiro13}
{Coleiro}, A. \& {Chaty}, S. 2013, \apj, 764, 185

\bibitem[{{Del Santo} {et~al.}(2007){Del Santo}, {Sidoli}, {Mereghetti},
  {Bazzano}, {Tarana}, \& {Ubertini}}]{delsanto07}
{Del Santo}, M., {Sidoli}, L., {Mereghetti}, S., {Bazzano}, A., {Tarana}, A.,
  \& {Ubertini}, P. 2007, \aap, 468, L17

\bibitem[{{Dubus} {et~al.}(2001){Dubus}, {Hameury}, \& {Lasota}}]{dubus01}
{Dubus}, G., {Hameury}, J.-M., \& {Lasota}, J.-P. 2001, \aap, 373, 251

\bibitem[{{Dubus} {et~al.}(1999){Dubus}, {Lasota}, {Hameury}, \&
  {Charles}}]{dubus99}
{Dubus}, G., {Lasota}, J.-P., {Hameury}, J.-M., \& {Charles}, P. 1999, \mnras,
  303, 139

\bibitem[{{Dutra} {et~al.}(2003){Dutra}, {Bica}, {Soares}, \&
  {Barbuy}}]{dutra03}
{Dutra}, C.~M., {Bica}, E., {Soares}, J., \& {Barbuy}, B. 2003, \aap, 400, 533

\bibitem[{{Forman} {et~al.}(1978){Forman}, {Jones}, {Cominsky}, {Julien},
  {Murray}, {Peters}, {Tananbaum}, \& {Giacconi}}]{forman78}
{Forman}, W., {Jones}, C., {Cominsky}, L., {Julien}, P., {Murray}, S.,
  {Peters}, G., {Tananbaum}, H., \& {Giacconi}, R. 1978, \apjs, 38, 357

\bibitem[{{Geminale} \& {Popowski}(2004)}]{geminale04}
{Geminale}, A. \& {Popowski}, P. 2004, AcA, 54, 375

\bibitem[{{Gilfanov}(2004)}]{gilfanov04}
{Gilfanov}, M. 2004, \mnras, 349, 146

\bibitem[{{Grimm} {et~al.}(2002){Grimm}, {Gilfanov}, \& {Sunyaev}}]{grimm02}
{Grimm}, H.-J., {Gilfanov}, M., \& {Sunyaev}, R. 2002, \aap, 391, 923

\bibitem[{{Grimm} {et~al.}(2003){Grimm}, {Gilfanov}, \& {Sunyaev}}]{grimm03}
{Grimm}, H.-J., {Gilfanov}, M., \& {Sunyaev}, R. 2003, \mnras, 339, 793

\bibitem[{{Hynes} {et~al.}(2004){Hynes}, {Charles}, {van Zyl}, {Barnes},
  {Steeghs}, {O'Brien}, \& {Casares}}]{hynes04}
{Hynes}, R.~I., {Charles}, P.~A., {van Zyl}, L., {Barnes}, A., {Steeghs}, D.,
  {O'Brien}, K., \& {Casares}, J. 2004, \mnras, 348, 100

\bibitem[{{Illarionov} \& {Sunyaev}(1975)}]{illarionov75}
{Illarionov}, A.~F. \& {Sunyaev}, R.~A. 1975, \aap, 39, 185

\bibitem[{{in 't Zand} {et~al.}(1998){in 't Zand}, {Bazzano}, {Cocchi},
  {Ubertini}, {Muller}, \& {Torroni}}]{zand98}
{in 't Zand}, J., {Bazzano}, A., {Cocchi}, M., {Ubertini}, P., {Muller}, J.~M.,
  \& {Torroni}, V. 1998, \iaucirc, 6846, 2

\bibitem[{{Jonker} \& {Nelemans}(2004)}]{jonker04}
{Jonker}, P.~G. \& {Nelemans}, G. 2004, \mnras, 354, 355

\bibitem[{{Krivonos} {et~al.}(2007){Krivonos}, {Revnivtsev}, {Lutovinov},
  {Sazonov}, {Churazov}, \& {Sunyaev}}]{krivonos07}
{Krivonos}, R., {Revnivtsev}, M., {Lutovinov}, A., {Sazonov}, S., {Churazov},
  E., \& {Sunyaev}, R. 2007, \aap, 475, 775

\bibitem[{{Krivonos} {et~al.}(2012){Krivonos}, {Tsygankov}, {Lutovinov},
  {Revnivtsev}, {Churazov}, \& {Sunyaev}}]{krivonos12}
{Krivonos}, R., {Tsygankov}, S., {Lutovinov}, A., {Revnivtsev}, M., {Churazov},
  E., \& {Sunyaev}, R. 2012, \aap, 545, A27

\bibitem[{{Landi} {et~al.}(2007){Landi}, {Masetti}, {Bassani}, {Capitanio},
  {Fiocchi}, {Bird}, {Hill}, {Gehrels}, {Markwardt}, \& {Perri}}]{landi07}
{Landi}, R., {et~al.} 2007, The Astronomer's Telegram, 1273, 1

\bibitem[{{Lasota}(2001)}]{lasota01}
{Lasota}, J.-P. 2001, NewAR, 45, 449

\bibitem[{{Lawrence} {et~al.}(2007){Lawrence}, {Warren}, {Almaini}, {Edge},
  {Hambly}, {Jameson}, {Lucas}, \& {Casali}}]{lawrence07}
{Lawrence}, A., {Warren}, S.~J., {Almaini}, O., {Edge}, A.~C., {Hambly}, N.~C.,
  {Jameson}, R.~F., {Lucas}, P., \& {Casali}, M. 2007, \mnras, 379, 1599

\bibitem[{{Li} {et~al.}(1977){Li}, {Lewin}, {Clark}, {Doty}, {Hoffman}, \&
  {Rappaport}}]{li77}
{Li}, F.~K., {Lewin}, W.~H.~G., {Clark}, G.~W., {Doty}, J., {Hoffman}, J.~A.,
  \& {Rappaport}, S.~A. 1977, \mnras, 179, 21P

\bibitem[{{Lin} {et~al.}(2012){Lin}, {Webb}, \& {Barret}}]{lin12}
{Lin}, D., {Webb}, N.~A., \& {Barret}, D. 2012, \apj, 756, 27

\bibitem[{{Liu} {et~al.}(2007){Liu}, {van Paradijs}, \& {van den
  Heuvel}}]{liu07}
{Liu}, Q.~Z., {van Paradijs}, J., \& {van den Heuvel}, E.~P.~J. 2007, \aap,
  469, 807

\bibitem[{{L{\"u}} {et~al.}(2012){L{\"u}}, {Zhu}, {Postnov}, {Yungelson},
  {Kuranov}, \& {Wang}}]{lu12}
{L{\"u}}, G.-L., {Zhu}, C.-H., {Postnov}, K.~A., {Yungelson}, L.~R., {Kuranov},
  A.~G., \& {Wang}, N. 2012, \mnras, 424, 2265

\bibitem[{{Lutovinov} {et~al.}(2005){Lutovinov}, {Revnivtsev}, {Molkov}, \&
  {Sunyaev}}]{lutovinov05}
{Lutovinov}, A., {Revnivtsev}, M., {Molkov}, S., \& {Sunyaev}, R. 2005, \aap,
  430, 997

\bibitem[{{Lutovinov} {et~al.}(2013){Lutovinov}, {Revnivtsev}, {Tsygankov}, \&
  {Krivonos}}]{lutovinov13}
{Lutovinov}, A.~A., {Revnivtsev}, M.~G., {Tsygankov}, S.~S., \& {Krivonos},
  R.~A. 2013, \mnras, 431, 327

\bibitem[{{Markwardt} \& {Swank}(2003)}]{markwardt03}
{Markwardt}, C.~B. \& {Swank}, J.~H. 2003, The Astronomer's Telegram, 156, 1

\bibitem[{{Marshall} {et~al.}(2006){Marshall}, {Robin}, {Reyl{\'e}},
  {Schultheis}, \& {Picaud}}]{marshall06}
{Marshall}, D.~J., {Robin}, A.~C., {Reyl{\'e}}, C., {Schultheis}, M., \&
  {Picaud}, S. 2006, \aap, 453, 635

\bibitem[{{Masetti} {et~al.}(2004){Masetti}, {Foschini}, {Palazzi}, {Beckmann},
  {Lund}, {Brandt}, {Westergaard}, {Amati}, {Caroli}, {Del Sordo}, {Di Cocco},
  {Durouchoux}, {Farinelli}, {Frontera}, {Orlandini}, \&
  {Zdziarski}}]{masetti04}
{Masetti}, N., {et~al.} 2004, \aap, 423, 651

\bibitem[{{Masetti} {et~al.}(2007){Masetti}, {Landi}, {Pretorius}, {Sguera},
  {Bird}, {Perri}, {Charles}, {Kennea}, {Malizia}, \& {Ubertini}}]{masetti07}
{Masetti}, N., {et~al.} 2007, \aap, 470, 331

\bibitem[{{Meyer} \& {Meyer-Hofmeister}(1981)}]{meyer81}
{Meyer}, F. \& {Meyer-Hofmeister}, E. 1981, \aap, 104, L10

\bibitem[{{Pickles}(1998)}]{pickles98}
{Pickles}, A.~J. 1998, \pasp, 110, 863

\bibitem[{{Polletta} {et~al.}(2007){Polletta}, {Tajer}, {Maraschi},
  {Trinchieri}, {Lonsdale}, {Chiappetti}, {Andreon}, {Pierre}, {Le F{\`e}vre},
  {Zamorani}, {Maccagni}, {Garcet}, {Surdej}, {Franceschini}, {Alloin},
  {Shupe}, {Surace}, {Fang}, {Rowan-Robinson}, {Smith}, \&
  {Tresse}}]{polletta07}
{Polletta}, M., {et~al.} 2007, \apj, 663, 81

\bibitem[{{Predehl} \& {Schmitt}(1995)}]{predehl95}
{Predehl}, P. \& {Schmitt}, J.~H.~M.~M. 1995, \aap, 293, 889

\bibitem[{{Ranalli} {et~al.}(2003){Ranalli}, {Comastri}, \&
  {Setti}}]{ranalli03}
{Ranalli}, P., {Comastri}, A., \& {Setti}, G. 2003, \aap, 399, 39

\bibitem[{{Ratti} {et~al.}(2010){Ratti}, {Bassa}, {Torres}, {Kuiper},
  {Miller-Jones}, \& {Jonker}}]{ratti10}
{Ratti}, E.~M., {Bassa}, C.~G., {Torres}, M.~A.~P., {Kuiper}, L.,
  {Miller-Jones}, J.~C.~A., \& {Jonker}, P.~G. 2010, \mnras, 408, 1866

\bibitem[{{Revnivtsev} {et~al.}(2011){Revnivtsev}, {Postnov}, {Kuranov}, \&
  {Ritter}}]{revnivtsev11}
{Revnivtsev}, M., {Postnov}, K., {Kuranov}, A., \& {Ritter}, H. 2011, \aap,
  526, A94

\bibitem[{{Revnivtsev} {et~al.}(2013){Revnivtsev}, {Kniazev}, {Karasev},
  {Berdnikov}, \& {Barway}}]{revnivtsev13}
{Revnivtsev}, M.~G., {Kniazev}, A., {Karasev}, D.~I., {Berdnikov}, L., \&
  {Barway}, S. 2013, Astronomy Letters, 39, 523

\bibitem[{{Revnivtsev} {et~al.}(2012){Revnivtsev}, {Zolotukhin}, \&
  {Meshcheryakov}}]{revnivtsev12}
{Revnivtsev}, M.~G., {Zolotukhin}, I.~Y., \& {Meshcheryakov}, A.~V. 2012,
  \mnras, 2513

\bibitem[{{Robitaille} {et~al.}(2008){Robitaille}, {Meade}, {Babler},
  {Whitney}, {Johnston}, {Indebetouw}, {Cohen}, {Povich}, {Sewilo}, {Benjamin},
  \& {Churchwell}}]{robitaille08}
{Robitaille}, T.~P., {et~al.} 2008, \aj, 136, 2413

\bibitem[{{Robitaille} {et~al.}(2007){Robitaille}, {Whitney}, {Indebetouw}, \&
  {Wood}}]{robitaille07}
{Robitaille}, T.~P., {Whitney}, B.~A., {Indebetouw}, R., \& {Wood}, K. 2007,
  \apjs, 169, 328

\bibitem[{{Russeil}(2003)}]{russeil03}
{Russeil}, D. 2003, \aap, 397, 133

\bibitem[{{Sakano} {et~al.}(2002){Sakano}, {Koyama}, {Murakami}, {Maeda}, \&
  {Yamauchi}}]{sakano02}
{Sakano}, M., {Koyama}, K., {Murakami}, H., {Maeda}, Y., \& {Yamauchi}, S.
  2002, \apjs, 138, 19

\bibitem[{{Sazonov} {et~al.}(2006){Sazonov}, {Revnivtsev}, {Gilfanov},
  {Churazov}, \& {Sunyaev}}]{sazonov06}
{Sazonov}, S., {Revnivtsev}, M., {Gilfanov}, M., {Churazov}, E., \& {Sunyaev},
  R. 2006, \aap, 450, 117

\bibitem[{{Schlafly} \& {Finkbeiner}(2011)}]{schlafly11}
{Schlafly}, E.~F. \& {Finkbeiner}, D.~P. 2011, \apj, 737, 103

\bibitem[{{Schlegel} {et~al.}(1998){Schlegel}, {Finkbeiner}, \&
  {Davis}}]{schlegel98}
{Schlegel}, D.~J., {Finkbeiner}, D.~P., \& {Davis}, M. 1998, \apj, 500, 525

\bibitem[{{Sidoli} \& {Mereghetti}(2003)}]{sidoli03}
{Sidoli}, L. \& {Mereghetti}, S. 2003, The Astronomer's Telegram, 147, 1

\bibitem[{{Swank} {et~al.}(1976){Swank}, {Becker}, {Pravdo}, \&
  {Serlemitsos}}]{swank76}
{Swank}, J.~H., {Becker}, R.~H., {Pravdo}, S.~H., \& {Serlemitsos}, P.~J. 1976,
  \iaucirc, 2963, 1

\bibitem[{{Taylor}(2005)}]{taylor05}
{Taylor}, M.~B. 2005, in Astronomical Society of the Pacific Conference Series,
  Vol. 347, Astronomical Data Analysis Software and Systems XIV, ed.
  {P.~Shopbell, M.~Britton, \& R.~Ebert}, 29--+

\bibitem[{{Thorstensen} {et~al.}(1980){Thorstensen}, {Bowyer}, \&
  {Charles}}]{thorstensen80}
{Thorstensen}, J.~R., {Bowyer}, S., \& {Charles}, P.~A. 1980, \apj, 238, 964

\bibitem[{{Tomsick} {et~al.}(2008){Tomsick}, {Chaty}, {Rodriguez}, {Walter}, \&
  {Kaaret}}]{tomsick08}
{Tomsick}, J.~A., {Chaty}, S., {Rodriguez}, J., {Walter}, R., \& {Kaaret}, P.
  2008, \apj, 685, 1143

\bibitem[{{Tomsick} {et~al.}(2009){Tomsick}, {Chaty}, {Rodriguez}, {Walter}, \&
  {Kaaret}}]{tomsick09}
{Tomsick}, J.~A., {Chaty}, S., {Rodriguez}, J., {Walter}, R., \& {Kaaret}, P.
  2009, \apj, 701, 811

\bibitem[{{Udalski}(2003)}]{udalski03}
{Udalski}, A. 2003, \apj, 590, 284

\bibitem[{{Wachter}(1997)}]{wachter97}
{Wachter}, S. 1997, \apj, 490, 401

\bibitem[{{Werner} {et~al.}(2004){Werner}, {in't Zand}, {Natalucci},
  {Markwardt}, {Cornelisse}, {Bazzano}, {Cocchi}, {Heise}, \&
  {Ubertini}}]{werner04}
{Werner}, N., {et~al.} 2004, \aap, 416, 311

\bibitem[{{Wijnands} {et~al.}(2006){Wijnands}, {Kuulkers}, {Muno}, {Cackett},
  {in't Zand}, {Maccarone}, {Fender}, {Grindlay}, {Homan}, {Rupen},
  {Cornelisse}, {Miller-Jones}, {van der Klis}, {Markwardt}, \&
  {Wang}}]{wijnands06b}
{Wijnands}, R., {et~al.} 2006, The Astronomer's Telegram, 892, 1

\bibitem[{{Wijnands} {et~al.}(2002){Wijnands}, {Miller}, \&
  {Wang}}]{wijnands02}
{Wijnands}, R., {Miller}, J.~M., \& {Wang}, Q.~D. 2002, \apj, 579, 422

\bibitem[{{Wright} {et~al.}(2010){Wright}, {Eisenhardt}, {Mainzer}, {Ressler},
  {Cutri}, {Jarrett}, {Kirkpatrick}, {Padgett}, {McMillan}, {Skrutskie},
  {Stanford}, {Cohen}, {Walker}, {Mather}, {Leisawitz}, {Gautier}, {McLean},
  {Benford}, {Lonsdale}, {Blain}, {Mendez}, {Irace}, {Duval}, {Liu}, {Royer},
  {Heinrichsen}, {Howard}, {Shannon}, {Kendall}, {Walsh}, {Larsen}, {Cardon},
  {Schick}, {Schwalm}, {Abid}, {Fabinsky}, {Naes}, \& {Tsai}}]{wright10}
{Wright}, E.~L., {et~al.} 2010, \aj, 140, 1868

\bibitem[{{Zolotukhin} \& {Revnivtsev}(2011)}]{zolotukhin10b}
{Zolotukhin}, I.~Y. \& {Revnivtsev}, M.~G. 2011, \mnras, 411, 620

\bibitem[{{Zolotukhin} {et~al.}(2010){Zolotukhin}, {Revnivtsev}, \&
  {Shakura}}]{zolotukhin10a}
{Zolotukhin}, I.~Y., {Revnivtsev}, M.~G., \& {Shakura}, N.~I. 2010, \mnras,
  401, L1

\end{thebibliography}

\label{lastpage}

\end{document}